# Toeplitz Inverse Eigenvalue Problem: Application to the Uniform Linear Antenna Array Calibration

Yuri Abramovich, *Life Fellow, IEEE*, Tanit Pongsiri

*Abstract*—The inverse Toeplitz eigenvalue problem (ToIEP) concerns finding a vector $r \in \mathbb{R}^N$ that specifies the real-valued symmetric Toeplitz matrix with the prescribed set of eigenvalues $(\lambda_1, \ldots, \lambda_N)$ [1]-[18]. Since phase "calibration" errors in uniform linear antenna arrays (ULAs) do not change the covariance matrix eigenvalues and the moduli of the covariance matrix elements, we formulate a number of the new ToIEP problems of the Hermitian Toeplitz matrix reconstruction, given the moduli of the matrix elements and the matrix eigenvalues. We demonstrate that for the real-valued case ($r \in \mathbb{R}^N$), only two solutions to this problem exist, with the "non-physical" one that in most practical cases could be easily disregarded. The computational algorithm for the real-valued case is quite simple. For the complex-valued case, we demonstrate that the family of solutions is broader and includes solutions inappropriate for calibration. For this reason, we modified the ToIEP problem to match the covariance matrix of the uncalibrated ULA. We investigate the statistical convergence of the algorithm with the sample matrices instead of the true ones. The proposed algorithms require the so-called "strong" or "argumental" convergence, which means a large enough required sample volume that reduces the errors in the estimated covariance matrix elements. Along with the ULA arrays, we also considered the fully augmentable minimum redundancy arrays that generate the same (full) set of covariance lags as the uniform linear arrays, and we specified the conditions when the ULA Toeplitz covariance matrix may be reconstructed given the $M$-variate MRA covariance matrix.

## I. INTRODUCTION

The broad class of the so-called "inverse eigenvalue problems" concerns reconstruction of a matrix from the prescribed eigendecomposition data [1]-[18]. These problems arise in a remarkable variety of applications and therefore, have very different formulations. In only one paper [13] does the author explicitly discuss 37 inverse eigenvalue problems, not counting many other implied variations. The forms and algorithms differ notably from problem to problem. The special class of "inverse eigenvalue problems" represents the "structured inverse eigenvalue problems" with an objective to construct a matrix that maintains both the specific structure, as well as the given spectral property [14].

For the uniform linear arrays (ULAs) or the "fully augmentable minimum redundancy arrays" (MRAs), we are interested in the reconstruction of the covariance matrix of the signals at the output of the ideally calibrated ULA. Correspondingly, we are interested in the class of the "Toeplitz inverse eigenvalue problems" (ToIEP) [6]. In [6], the ToIEP problem is formulated for the real-valued symmetric Toeplitz matrices when "an inverse eigenvalue problem" (ToIEP) concerns finding a vector $r \in \mathbb{R}^N$ so that $\mathbf{T}(r)$ has a prescribed set of real numbers $(\lambda_1, \ldots, \lambda_N)$ as its eigenvalues. Here $r$ is the first row of the $N$-variate symmetric Toeplitz matrix.

For signals with the symmetric spatial spectrum, this ToIEP methodology is directly appropriate, though the problem of the symmetric Toeplitz matrix reconstruction with the given moduli of the matrix's elements has not been yet addressed. We obviously had to explore the more general case, where $r \in \mathbb{C}^N$ describes the complex-valued Hermitian Toeplitz covariance matrix. We selected the ToIEP theory "toolbox" due to the fact that the phase "calibration" errors do not affect eigenvalues of the covariance matrix and the moduli of the covariance matrix elements. Therefore, if the reconstruction of a Toeplitz covariance matrix, given its eigenvalues and moduli of the matrix's elements are unique, the ULA calibration problem would be solved.

Yet, it is clear that for any Hermitian Toeplitz matrix $\mathbf{T}_N$, the same eigenvalues and moduli of the covariance matrix elements have the matrices within the family of Hermitian Toeplitz matrices $\mathbf{J}_N$:

$$\mathbf{J}_N := \mathbf{D}(\boldsymbol{\varphi})\mathbf{T}_N\mathbf{D}(\boldsymbol{\varphi}) \qquad (1)$$

where

$$\mathbf{D}(\boldsymbol{\varphi}) := \text{diag}\left[1, \exp(j\boldsymbol{\varphi}), \ldots, \exp(j(N-1)\boldsymbol{\varphi})\right] \qquad (2)$$

For the real-valued case, only

$$\varphi = \pi \qquad (3)$$

in (2) retains the real-valued symmetric Toeplitz matrix within the same class. Therefore, if the uniqueness of the Toeplitz reconstruction problem is impossible in our problem, then the main question is whether (1), (2) are the unique solutions family. Uniqueness of this family means that for the Toeplitz covariance matrix reconstruction we restore the original manifold of the ideal ULA, removing all random-like errors. The remaining unspecified shift $\mathbf{D}(\varphi_N)$ means that the calibrated array has an arbitrary bias that can be removed by using a weak source with known coordinates.

This study is inspired by the calibration problem of the long-aperture receive antenna arrays of the modern HF Over-the-Horizon radars (HF OTHR), where a weak stellar source, such as the radio star Cygnus A, may be used for bias removal

Yuri Abramovich and Tanit Pongsiri are with WR Systems, Ltd., Fairfax, VA 22030, USA. Emails: yabramovich@wrsystems.com, tpongsiri@wrsystems.com

[30]-[32]. In fact, most of the proposed ULA calibration techniques require this additional step of bias removal [21], [22].

In terms of uniqueness of the solutions family (1), (2) in our study, we demonstrate that the complex-valued and the real-valued cases are very different in this respect. Specifically, in Sec III we demonstrate two examples where one fits the family (1), (2) and the other does not. This forced us to further specify the problem of the complex-valued Hermitian covariance matrix reconstruction fitting the given Hermitian covariance matrix of the uncalibrated ULA in more detail. For the real-valued symmetric Toeplitz matrix, the situation is very different. Both existing theoretical results [4] and the conducted reconstructions suggest that for this case, only two solutions mentioned above (1)-(3) exist for the problem of the symmetric Toeplitz matrix reconstruction, given the elements moduli and eigenvalues.

While the difference between the complex-valued and real-valued cases is somewhat disappointing, it is worth mentioning that in HF OTHR the strongest returns are from the Earth's surface. These returns are localized in range and Doppler frequency and while being illuminated by the transmitting antenna with a symmetric main beam pattern, these returns have a symmetric spatial spectrum (in $\sin\theta$ coordinates). Naturally, these powerful clutter returns, illuminated by a cooperative or non-cooperative HF OTHR transmitter, may be used for the receive array calibration. The selection of the appropriate (one of the two) solutions for calibration is quite straightforward in most cases, as demonstrated below.

In accordance with the purpose of this paper, in Sec II we provide the accurate analytical description of the phase calibration problem and the associated ToIEPs. In Sec III, we introduce the reconstruction method we applied for the Hermitian covariance matrix reconstruction and then the two examples mentioned above, one of which belongs to the family (1), (2), while the other one does not. This motivated us to reformulate the reconstruction problem by searching for the Hermitian matrix with the given elements moduli and eigenvalues, such that the Hadamard (element-wise) division of the given Hermitian matrix by the reconstructed Toeplitz matrix is a rank-one matrix.

In Sec IV, we introduce our reconstruction technique for the real-valued Toeplitz matrix and provide the associated theoretical results regarding uniqueness of the solution. In Sec V, we introduce the results of the Monte-Carlo simulations of the Toeplitz matrix reconstruction for the real-valued ULA and MRA covariance matrix cases. In Sec VI, we specify a technique for selection of one of the two reconstructions that fits our antenna calibration problem. In Sec VII, we demonstrate the "strong convergence" requirement for the problem with the sample matrices replacing the true covariance matrices. This "strong convergence" requirement means the sample volume should be large enough to make the estimation errors in the covariance matrices elements minimal. In Sec VIII, we conclude our paper.

## II. Problem Formulation

Consider a uniform linear $N$-element array (ULA) with uniform mutual coupling and equalized antenna element gains.

In the absence of phase "calibration" errors, the manifold of this truly uniform array is described by the ideal ULA model:

$$\mathbf{a}_N^{\mathrm{T}}(\theta) = \left[1, \exp\left(j\frac{2\pi}{\lambda}d\sin\theta\right), \dots, \exp\left(j\frac{2\pi}{\lambda}d(N-1)\sin\theta\right)\right], \quad (4)$$

where:

$d$ is the inter-element spacing
$\lambda$ is the wavelength
$\theta$ is the coning angle (azimuth) calculated off the antenna boresight

For the ULA with the uncalibrated phase errors $\varphi_j$, $j = 0,1,\dots N-1$ ($\varphi_o = 0^o$), the modified (by errors) manifold vector $\mathbf{C}_N$ is

$$\mathbf{C}_N = \mathbf{D}_N(\boldsymbol{\Phi}_{N-1})\mathbf{A}_N(\theta) \quad (5)$$

where

$$\mathbf{D}_N(\boldsymbol{\Phi}_{N-1}) = \mathrm{diag}\left[1, \exp(j\varphi_1), \dots, \exp(j\varphi_{N-1})\right] \quad (6)$$

Since the initial phase $\varphi_o$ does not affect the covariance matrix, we assumed $\varphi_o = 0^o$ (and $\varphi_n' - \varphi_o' = \varphi_n$) and $\varphi_j$ is distributed over $[-\pi, \pi]$. In this study, the particular phase errors distribution is irrelevant. The truly random-like phase errors should have a zero linear trend across the array

$$\sum_{n=1}^{N-1}\frac{\varphi_n}{n} = 0 \quad (7)$$

since any linear trend in phase errors is causing the above mentioned bias in azimuth $\theta$ estimation, which needs to be removed.

For the beampattern of the embedded ULA element $p(\theta)$ and the angular distribution of the signal arriving from the front hemisphere $[|\theta| \leq \pi/2]$, the spatial covariance matrix at the uncalibrated ULA output is:

$$\mathbf{R}_N[f(\theta)] = \mathbf{D}_N[\boldsymbol{\Phi}_{N-1}]\mathbf{T}_N\mathbf{D}_N^{\mathrm{H}}[\boldsymbol{\Phi}_{N-1}] \quad (8)$$

where

$$\mathbf{T}_N[f(\theta)] = \int_{-\pi/2}^{\pi/2} p(\theta)f(\theta)\mathbf{A}_N(\theta)\mathbf{A}_N^{\mathrm{H}}(\theta)d\theta \quad (9)$$

is the Hermitian Toeplitz covariance matrix at the output of the ideal (phase errors-free) ULA array; $f(\theta)$ is the spatial distribution of the received signals. The unitary nature of the transformation from $\mathbf{T}_N$ (9) to $\mathbf{R}_N$ (8), with

$$\mathbf{D}_N(\boldsymbol{\Phi}_{N-1})\mathbf{D}_N^{\mathrm{H}}(\boldsymbol{\Phi}_{N-1}) = \mathbf{D}_N^{\mathrm{H}}(\boldsymbol{\Phi}_{N-1})\mathbf{D}_N(\boldsymbol{\Phi}_N) = \mathbf{I}_N \quad (10)$$

leads to the two important properties that are actually "responsible" for invoking the "structured inverse eigenvalue problem" toolbox.

<u>Property 1</u>.     Eigenvalues of the covariance matrix $\mathbf{R}_N$ of the output signals for the uncalibrated ULA array are exactly the same as the eigenvalues of the ideally calibrated ULA covariance matrix.

<u>Property 2</u>.     Moduli of the covariance matrix $\mathbf{R}_N$ elements $|r_{pl}|$ are exactly the same as the moduli $|t_{p-l}|$ of the Toeplitz covariance matrix of the ideally calibrated ULA:

$$r_{pl} = t_{p-l} \exp[j(\varphi_p - \varphi_l)] \tag{11}$$

Correspondingly, the new "Toeplitz inverse eigenvalue" problem may be formulated as follows.

<u>Problem A</u>.     Given all $N$ non-negative eigenvalues of the p.d. Toeplitz Hermitian (covariance) matrix and moduli of all elements of this matrix, reconstruct a Toeplitz matrix that belongs to the family (8), (9) of matrices with the same elements moduli and eigenvalues.

This family $\mathbf{J}_N$ is defined in (8), (9) as:

$$\mathbf{J}_N := \mathbf{D}_N(\boldsymbol{\psi}_{N-1})\mathbf{T}_N\mathbf{D}_N^H(\boldsymbol{\psi}_{N-1}) \tag{12}$$

$$\mathbf{D}_N(\boldsymbol{\psi}_{N-1}) = \operatorname{diag}\left[1, \exp(j\boldsymbol{\psi}), \dots, \exp(j\,(N-1)\boldsymbol{\psi})\right] \tag{13}$$

As discussed above, it is clear that if the family (12), (13) is the only family of solutions for problem A, then the reconstruction of any matrix within this family will mean the restoration of the ideal ULA manifold with the need for additional bias mitigation. While mathematically proving the uniqueness of this solutions family (12), (13) is a complicated problem, a single counter-example is sufficient to prove this hypothesis wrong. Such an example is provided in Sec III.

This proven non-uniqueness of the family (12), (13) for the complex-valued case of the Hermitian matrix reconstruction made us add an additional condition in Problem A and formulate Problem B.

<u>Problem B</u>.     Given $N$-variate Hermitian (covariance) p.d. matrix, reconstruct a Toeplitz Hermitian matrix with the same eigenvalues and moduli of the matrix elements, such that the element-wise (Hadamard) division of the given Hermitian matrix by the reconstructed Toeplitz matrix is the rank-one matrix with a constant modulus eigenvector (for the non-zero eigenvalue).

In Sec III we investigate whether this additional condition allows for the solutions to belong to the family (12), (13).

As previously mentioned, it is quite a different situation when the reconstructed matrix is a real-valued symmetric Toeplitz covariance matrix. Let us first specify the class of the practical problems described by the real-valued symmetric

Toeplitz covariance matrices. Let us consider the symmetric angular distribution $f(\sin\theta)$ in (9)

$$f(\sin\theta) = f[|\sin\theta - \sin\theta_o| \le \Delta] \tag{14}$$

In Sec IV for our numerical examples we consider the rectangular distribution, symmetric with respect to its center:

$$f(\sin\theta) = \begin{cases} 1, & \text{for } \sin\theta_{min} < \sin\theta < \sin\theta_{max} \\ \frac{1}{2}, & \text{for } \sin\theta = \begin{cases} \sin\theta_{min} \\ \sin\theta_{max} \end{cases} \\ 0, & \text{elsewhere} \end{cases} \tag{15}$$

$$f(\sin\theta) = \operatorname{rect}\left[\sin\theta - \sin\theta_o\right] \tag{16}$$

where

$$\sin\theta_o = \frac{\sin\theta_{max} + \sin\theta_{min}}{2}$$

$$\Delta = \frac{\sin\theta_{max} - \sin\theta_{min}}{2}$$

(for $\sin\theta_{max}, \sin\theta_{min} > 0$)

For the beampattern of the embedded antenna element $p(\theta)$:

$$p(\theta) = \cos\theta, \tag{17}$$

the element $t_{p-l}$ of the ideal (error-free) Toeplitz covariance matrix $\mathbf{T}_N[f(\theta)]$ is therefore defined as

$$t_{p-l} = \int_{\sin\theta_o - \Delta}^{\sin\theta_o + \Delta} f[|\sin\theta - \sin\theta_o| \\ \times \exp[j\frac{2\pi d}{\lambda}(p-l)\sin\theta] \ d\theta \tag{18}$$

For

$$x = \sin\theta - \sin\theta_o, \quad dx = \cos\theta d\theta \tag{19}$$

we have,

$$t_{p-l} = \exp\left[j\frac{2\pi d}{\lambda}(p-l)\sin\theta_o\right] \\ \times \int_{-\Delta}^{\Delta} f(x)\exp\left[j\frac{2\pi d}{\lambda}(p-l)x\right]dx \tag{20}$$

Since

$$2\int_o^{\Delta}\cos\left[\frac{2\pi d}{\lambda}(p-l)x\right]dx \\ = \frac{\lambda}{d}\frac{\sin[\frac{2\pi d}{\lambda}(p-l)\Delta]}{\pi(p-l)} \tag{21}$$

we have:

$$\mathbf{T}_N = \mathbf{D}_N\left[\frac{2\pi d}{\lambda}\sin\theta_o\right]\mathbf{T}_{NR}(W)\,\mathbf{D}_N^H\left[\frac{2\pi d}{\lambda}\sin\theta_o\right] \tag{22}$$

where

$$\mathbf{T}_{NR} = \frac{\lambda}{d}\, \text{sinc}_N(W) \tag{23}$$

$$\text{sinc}_N(W) = \frac{\sin\,[2\pi W\,(p-l)]}{\pi(p-l)} \tag{24}$$

$$W = \frac{d}{\lambda}\frac{\sin\theta_{max} - \sin\theta_{min}}{2} \tag{25}$$

$$\mathbf{D}_N\left[\frac{2\pi d}{\lambda}\sin\theta\right] = \text{diag}\left[\exp\left(j\frac{2\pi d}{\lambda}n\sin\theta\right)\right] \tag{26}$$
$$n = 0,\dots,N-1$$

One can see that the real-valued Toeplitz matrix $\mathbf{T}_{NR}(W)$ is the covariance matrix of the radar return illuminated by the rectangular beampattern of the Tx antenna array (in $\sin\theta$ coordinates), shifted by the matrix $\mathbf{D}_N[(2\pi d/\lambda)\sin\theta_o]$ into the direction $\sin\theta_o$. Therefore, in the presence of random errors the Hermitian covariance matrix $\mathbf{R}_N$ is

$$\mathbf{R}_N[f(\theta)] = \mathbf{D}_N[\mathbf{\Phi}_{N-1}]\mathbf{D}_N\left[\frac{2\pi d}{\lambda}\sin\theta_o\right]\mathbf{T}_{NR}$$
$$\times \mathbf{D}_N^{\mathrm{H}}\left(\frac{2\pi d}{\lambda}\sin\theta_o\right)\mathbf{D}_N[\mathbf{\Phi}_{N-1}] \tag{27}$$

In Sec IV, apart from the rectangular spectrum, we also analyze the case of the exponential angular spectrum:

$$f(x) = \exp(-\alpha|x|)\,,\quad |x| < \Delta \tag{28}$$

when

$$t_{p-l}$$
$$= \frac{e^{-2\alpha W}[\pi(p-l)\sin(2\pi W(p-l) - \alpha\cos(2\pi W(p-l))] + \alpha}{\alpha^2 + [\pi(p-l)]^2} \tag{29}$$

For $\alpha = 0$, we have

$$t_{p-l} = \frac{\sin[2\pi W(p-l)]}{\pi(p-l)} \qquad \text{(see (24))}$$

For $\alpha \neq 0, p - l = 0$, we have

$$t_o = \frac{1 - e^{-2\alpha W}}{\alpha} \tag{30}$$

so that for $\alpha = 0$, when $\exp(-2\alpha W) \to 1 - 2\alpha W$, $t_o$ tends to

$$t_o = 2W \tag{31}$$

Therefore, for the described problems with the symmetric spatial spectrum, the antenna calibration problem may be addressed by the reconstruction of the real-valued Toeplitz matrix $\mathbf{T}_{NR}$, when the eigenvalues and the elements moduli of the reconstructed real-valued matrix are the same as per the given complex-valued covariance matrix $\mathbf{R}_N$. This reconstruction should "remove" the random-like phase errors, while the bias caused by the lost angle shift produced by the diagonal matrix $\mathbf{D}_N[(2\pi d/\lambda)\sin\theta_o]$ should be restored by an external (weak) source with a known azimuth or by the original beamforming, if the phase errors are moderate. Naturally, this could be done only if the solutions family (8), (9) is unique for this reconstruction problem. Therefore, from the "structured inverse eigenvalue problem" standpoint, the following problem has to be addressed.

<u>Problem C.</u>  Given all $N$ non-negative eigenvalues of the real-valued symmetric Toeplitz matrix and all moduli of these matrix elements, reconstruct the symmetric Toeplitz matrix with these eigenvalues and elements moduli.

It is clear that for this real-valued case, the proper sign inversions should be administered to the sub-diagonals of the matrix built with the given moduli of the matrix elements. The positive outcome of the problem C solution should encourage addressing the problem of a minimum redundancy sub-array (MRA) calibration, given a test signal with a symmetric spatial spectrum.

Let us introduce the $(M \times N)$–variate selection matrix in every row with a single non-zero element equal to one in the position of the MRA $j$-th element within the $N$-variate ULA's aperture. Then, the covariance matrix at the output of the $M$-variate MRA is:

$$\mathbf{R}_M[f(\theta)]$$
$$= \mathbf{H}_{M,N}[\mathbf{D}_N[\mathbf{\Phi}_{N-1}]\mathbf{T}_N\mathbf{D}_N^{\mathrm{H}}[\mathbf{\Phi}_{N-1}]]\mathbf{H}_{M,N}^{\mathrm{T}} \tag{32}$$

It is straightforward to show that

$$\mathbf{H}_{M,N}\mathbf{D}[\mathbf{\Phi}_{N-1}] = \mathbf{D}_M[\mathbf{\Phi}_{M-1}]\mathbf{H}_{M,N} \tag{33}$$

where $\mathbf{D}_M[\mathbf{\Phi}_{M-1}]$ is the $M$-variate diagonal matrix with the elements $\exp(j\varphi_m)$, where $m = 0,\dots,M-1$, $\varphi_o = 0$, on the main diagonal. Therefore, we have:

$$\mathbf{R}_M[f(\theta)]$$
$$= \mathbf{D}_M[\mathbf{\Phi}_{M-1}]\big[\mathbf{H}_{M,N}\mathbf{T}_{NR}(f(\theta))\mathbf{H}_{M,N}^{\mathrm{T}}\big]\mathbf{D}_M^{\mathrm{H}}[\mathbf{\Phi}_{M-1}] \tag{34}$$

with

$$\mathbf{H}_{M,N}\mathbf{T}_N(f(\theta))\mathbf{H}_{M,N}^{\mathrm{T}} = \mathbf{R}_M^o[f(\theta)] \tag{35}$$

where $\mathbf{R}_M^o[f(\theta)]$ is the MRA covariance matrix of the phase errors-free MRA array.

From (35) it once again follows that the eigenvalues of the matrices $\mathbf{R}_M[f(\theta)]$ and $\mathbf{R}_M^o[f(\theta)]$ are the same. The moduli of the matrices elements are the same as well. The latter means that the moduli of the matrix $\mathbf{R}_M[f(\theta)]$ elements specify the moduli of the Toeplitz $N \times N$-variate matrix, if the MRA is fully augmentable. Therefore, given the moduli of all $N$ elements of the Toeplitz symmetric matrix and $M$ eigenvalues of the MRA transformation (32), we may try to reconstruct the entire real-valued Toeplitz matrix. Obviously, the missing $(N - M)$ eigenvalues may not allow us to solve this problem unambiguously. The special case with the signal subspace dimension, which is smaller than the MRA dimension, may provide the $(N - M)$ missing noise subspace eigenvalues of the $N$-variate Toeplitz matrix and allow for the successful problem solution. We explore these options in Sec. IV.

## III. RECONSTRUCTION OF THE COMPLEX-VALUED HERMITIAN TOEPLITZ COVARIANCE MATRIX, GIVEN ITS EIGENVALUES AND MODULI OF THE MATRIX ELEMENTS

As discussed in Introduction, an example with two Hermitian Toeplitz matrices with the same eigenvalues and moduli of the matrices elements, not linked by the transformation (12) describing the desired family of solutions, is sufficient to prove the hypothesis wrong. The existence of such an example, regardless of the way it was derived, is sufficient to treat the condition of the matrix elements moduli and eigenvalues as not belonging to the family (12). Yet, since the reconstruction of a Hermitian Toeplitz matrix with the given moduli of the matrix elements may be used in other applications, below we introduce the numerical technique these examples were derived from.

In [1], it was proposed for the affine family of real-valued matrices:

$$\mathbf{A}(x) = \mathbf{A}_o + \sum_{k=1}^{N} x_K \mathbf{A}_K \ , \tag{36}$$

where $x \in \mathbb{R}^N$ and $\{\mathbf{A}_K\}$ are the real symmetric $N \ x \ N$ matrices. The method is applied for solving the equation

$$f(x) = \begin{bmatrix} \lambda_1(x) - \lambda_1^* \\ \vdots \\ \lambda_N(x) - \lambda_N^* \end{bmatrix} = 0 \tag{37}$$

where $\lambda_1^*, \dots, \lambda_N^*$ are the given eigenvalues. For the distinct given eigenvalue $\lambda_j^*, j = 1, \dots, N$ and differentiable in a neighborhood of $\mathbf{x}_{opt}$ eigenvalues, the authors in [1] proved the quadratic convergence of the Newton's algorithm:

$$\mathbf{J}(\mathbf{x}^{(n)})(\mathbf{x}^{(n+1)} - \mathbf{x}^{(n)}) = -f(\mathbf{x}^{(n)}) \tag{38}$$

where

$$\mathbf{J}_{jk} = \frac{\partial \lambda_j}{\partial x_k} = \mathbf{U}_j^{\mathrm{H}}(x)\mathbf{A}_K\mathbf{U}_j(x) \ , \tag{39}$$

and $\mathbf{U}_j(x), j = 1, \dots, N$ are the eigenvectors of the matrix $\mathbf{A}(x)$. Analysis of the algorithm (38) for the multiple (equal) eigenvalues revealed some problems with its convergence. Yet, the authors commented in [1] that "even when no modifications are made to method (38), (39), it is locally quadratically convergent regardless of eigenvalues multiplicity, assuming the problem has a solution".

The authors in [1] also suggested the modification to the algorithm (38), (39), when $\mathbf{A}(x)$ is not affine, but is a nonlinear function of $x$. They suggested in [1] that (39) should be replaced by:

$$\frac{\partial \lambda_j(x)}{\partial x_k} = \mathbf{U}_j^{\mathrm{H}}(x)\mathbf{A}_k(x)\mathbf{U}_j(x) \ , \tag{40}$$

where

$$\mathbf{A}_K(x) = \frac{\partial \mathbf{A}(x)}{\partial x_k} \tag{41}$$

Using the recommendations provided in [1], we presented the optimized Hermitian Toeplitz matrix as:

$$\mathbf{T}_N(\boldsymbol{\psi}_{N-1}) = t_o\mathbf{I}_N + \sum_{k=1}^{N-1} |t_k|\mathbf{A}_k(\psi_k) \tag{42}$$

where $t_o, |t_k|, k = 1, \dots, N-1$ are the provided moduli, and

$$
\begin{aligned}
&\mathbf{A}_k(\psi_k) \\
&= \begin{bmatrix} 0 & \dots & 0 & e^{j\psi_k} & 0 & & 0 \\ \vdots & & & & e^{j\psi_k} & & \\ 0 & & & & & & \\ e^{-j\psi_k} & & & & & \ddots & \\ 0 & e^{-j\psi_k} & & & & & e^{j\psi_k} \\ & & \ddots & & & & \\ 0 & \dots & 0 & e^{-j\psi_k} & 0 & \dots & 0 \end{bmatrix}
\end{aligned} \tag{43}
$$

The non-linear equations in (37) are:

$$\begin{bmatrix} \lambda_1(\boldsymbol{\psi}_{N-1}) - \lambda_1^* \\ \vdots \\ \lambda_N(\boldsymbol{\psi}_{N-1}) - \lambda_N^* \end{bmatrix} = 0 \tag{44}$$

The eigenvalue of $\lambda_j(\boldsymbol{\psi}_{N-1})$ is therefore equal to:

$$
\begin{aligned}
&\lambda_j(\boldsymbol{\psi}_{N-1}) \\
&= t_o + \sum_{k=1}^{N-1} |t_k|\mathbf{U}_j^{\mathrm{H}}(\boldsymbol{\psi}_{N-1})\mathbf{A}_K(\boldsymbol{\psi}_{N-1})\mathbf{U}_j(\boldsymbol{\psi}_{N-1})
\end{aligned} \tag{45}
$$

Differentiating the relations [10]:

$$\begin{cases} \mathbf{U}_j^{\mathrm{H}}(\boldsymbol{\psi}_{N-1})\mathbf{U}_j(\boldsymbol{\psi}_{n-1}) = 1 \\ \mathbf{U}_j^{\mathrm{H}}(\boldsymbol{\psi}_{N-1})\mathbf{T}_N(\boldsymbol{\psi}_{N-1})\mathbf{U}_j(\boldsymbol{\psi}_{N-1}) = \lambda_j \end{cases} \tag{46}$$

similar to (2.28) in [1], [10], we have:

$$
\begin{aligned}
&\frac{\partial \lambda_j(\boldsymbol{\psi}_{N-1})}{\partial \psi_K} \\
&= \mathbf{U}_j^{\mathrm{H}}(\boldsymbol{\psi}_{N-1})\mathbf{B}_K(\boldsymbol{\psi}_{N-1})\mathbf{U}_j(\boldsymbol{\psi}_{N-1})|t_k| \ ,
\end{aligned} \tag{47}
$$

where

$$\mathbf{B}_K = \frac{\partial \mathbf{A}_K(\psi_k)}{\partial \psi_k} \tag{48}$$

$$\mathbf{B}_K(\psi_k) \tag{49}$$

$$= \begin{bmatrix} 0 & \dots & 0 & je^{j\psi_k} & \dots & 0 \\ \vdots & & & & \ddots & \\ -je^{-j\psi_k} & & & & & 0 \\ 0 & \ddots & & & & je^{j\psi_k} \\ & & \ddots & & & \\ 0 & \dots & 0 & -je^{-j\psi_k} & \dots & 0 \end{bmatrix}$$

Thus, the Jacobian of $f(\boldsymbol{\psi}_{N-1})$ is

$$\mathbf{J}_{jk} = |t_k| \mathbf{U}_j^{\mathrm{H}}(\boldsymbol{\psi}_{N-1}) \mathbf{B}_k(\psi_k) \mathbf{U}_j(\boldsymbol{\psi}_{N-1}) \ , \tag{50}$$

and the one step of the Newton's method is defined as:

$$\mathbf{J}(\boldsymbol{\psi}_{N-1}^{(t)})[\boldsymbol{\psi}_{N-1}^{(t+1)} - \boldsymbol{\psi}_{N-1}^{(t)}] = -f[\boldsymbol{\psi}_{N-1}^{(t)}] \tag{51}$$

While the expression (51) looks the same as the expression (2.6) in [1], it needs some specifications. First, note that $\mathbf{J}(\boldsymbol{\psi}_{N-1}) \in \mathbb{C}^{N \times N-1}$, i.e., $\mathbf{J}(\boldsymbol{\psi}_{N-1})$ is a "tall" matrix with $N$ rows and $N$ - 1 columns. Yet, since the main diagonal of the reconstructed Toeplitz matrix (42) remains constant, and since the sum of all eigenvalues of the matrix $\mathbf{T}_N$ in (42) is equal to the matrix's trace:

$$\sum_{j=1}^{N} \lambda_j(\boldsymbol{\psi}_{N-1}) = t_o N \ , \tag{52}$$

The number of "free" eigenvalues in the matrix (42) is also equal to $N - 1$. This allows for reducing the vector (51) to $(N - 1)$-dimensions and keeping the Jacobian (50) as a square matrix with $j=2,\dots,N$ and $k=1,\dots,N$ - 1. Therefore, for the non-degenerate $\mathbf{J}(\boldsymbol{\psi}_{N-1}^{(t)})$ in (51), we have

$$\boldsymbol{\psi}_{N-1}^{(t+1)} = \boldsymbol{\psi}_{N-1}^{(t)} - \mathbf{J}^{-1}(\boldsymbol{\psi}_{N-1}^{(t)}) f_{N-1}(\boldsymbol{\psi}_{N-1}^{(t)}) \tag{53}$$

Second, the point $\boldsymbol{\psi}_{N-1} = 0$ is not appropriate for the initialization of this iterative process. Indeed, for $\boldsymbol{\psi}_{N-1} = 0$, we have:

$$\frac{\partial \lambda_j(0)}{\partial \psi_k} = |t_k| \mathbf{U}_j^{\mathrm{H}}(0) \mathbf{B}_k(0) \mathbf{U}_j(0). \tag{54}$$

In this case

$$\mathbf{B}_k(0) = j[\mathbf{J}(1)^k - (\mathbf{J}(-1)^k)] \tag{55}$$

where $\mathbf{J}(1)$ is the matrix with the first upper sub-diagonal with all the elements equal to 1. $\mathbf{J}(-1)$ is the similar matrix with the first lower sub-diagonal elements equal to 1. All other elements in both matrices are equal to zero. Note that the eigenvectors of a Hermitian Toeplitz matrix are the Hermitian vectors, i.e.:

$$\mathbf{J}\mathbf{x} = \mathbf{x}^* \tag{56}$$

For $\mathbf{T}_N(0)$, constructed of all positive moduli, we have:

$$\mathbf{J}\mathbf{x} = \mathbf{x}, \tag{57}$$

since $\mathbf{x} \in \mathbb{R}^N$, and therefore

$$\mathbf{x}^{\mathrm{T}}[\mathbf{J}^{(1)}]^k \mathbf{x} = \mathbf{x}^{\mathrm{T}}[\mathbf{J}^{(-1)}]^k \mathbf{x} \tag{58}$$

and therefore $\boldsymbol{\psi}_{N-1}^{(0)} = 0$ is not an admissible point for initialization of the Newton's algorithm (53).

Third, and most importantly, the algorithm (53) may work for phases $\boldsymbol{\psi}_{N-1}$ optimization only over small enough innovation $|\Delta(\boldsymbol{\psi}_{N-1}^{(t)})|$, where

$$\Delta(\boldsymbol{\psi}_{N-1}^{(t+1)}) = \mathbf{J}^{\mathrm{H}}(\boldsymbol{\psi}_{N-1}^{(t)}) f(\boldsymbol{\psi}_{N-1}^{(t)}) \tag{59}$$

and

$$\exp(j\psi_{N-1}^{(t+1)}) = \exp(j\psi_N^{(t)})(1 + j\Delta(\boldsymbol{\psi}_{N-1}^{(t+1)})) \tag{60}$$

which locally retains the affine approximations.

To secure operations in this locally affine approximation regime, we modified the algorithm (53)

$$\boldsymbol{\psi}_{N-1}^{(t+1)} = \boldsymbol{\psi}_{N-1}^{(t)} - c_{t+1}\Delta(\boldsymbol{\psi}_{N-1}^{(t+1)}) \tag{61}$$

so that

$$\max_k |c_{t+1}\Delta(\boldsymbol{\psi}_{N-1}^{(t+1)})| < 15^o \tag{62}$$

Usually, $|\Delta\boldsymbol{\psi}_{N-1}^{(t+1)}|$ is maximal over several initial iterations where we need to apply the scaling factor $c_{t+1}$. Obviously, with respect to these modifications, no claims on the type of convergence of the algorithm (61)-(62) could be made. Yet, our purpose was to get a solution, while the optimization of the algorithm to improve its convergence rate is an outstanding problem. In fact, by using the algorithm (59)-(62) we were able to converge to $|f(\boldsymbol{\psi}_{N-1}^{(t+1)})| < 10^{-6}$ in most cases. For this reason, we continued our optimization by the alternating projections that brought us to $|f(\boldsymbol{\psi}_{N-1}^{(t+1)})| < 10^{-9} - 10^{-12}$.

Specifically, for the Toeplitz matrix derived by the algorithm (59)-(62) with the given elements moduli, we found its eigendecomposition and reconstructed the new matrix with the derived eigenvectors and prescribed eigenvalues. Since this reconstructed matrix is not necessarily a Toeplitz matrix, we applied redundancy averaging and continued iterations. After a number of these alternating projections, we get a better match for the reconstructed Toeplitz matrix eigenvalues with the prescribed set, but the elements moduli of the reconstructed matrix may not be precisely equal to the prescribed values.

For the numerical trials that provided us with the examples of interest, we selected the $N$-variate ($N = 20$) Toeplitz matrix, which is the sum of two sinc($W$) matrices:

$$\begin{aligned} \mathbf{T}_N &= \mathrm{sinc}\,(W_1) \\ &+ 0.5\,\mathrm{diag}\,(\mu)\,\mathrm{sinc}\,(W_2)\mathrm{diag}^{\mathrm{H}}(\mu), \end{aligned} \tag{63}$$

where

$$\text{sinc}(W) = \left[ \frac{\sin 2\pi W(p-l)}{\pi(p-l)} \right] \quad p, l = 1, \ldots, 20 \quad (64)$$

and

$$\text{diag}(\mu) = \text{diag}[1, \exp\left( j\frac{2\pi d}{\lambda} \sin\theta_o \right), \ldots,$$
$$\exp\left( j(N-1)\frac{2\pi d}{\lambda} \sin\theta_o \right)] \quad (65)$$

with $W_1$ and $W_2$ different in our examples. We start from the successful example with $W_1 = 0.2$, $W_2 = 0.1$, $\theta_o = 20^o$. By the Newton's algorithm we achieved $f(\psi_{N-1}) < 10^{-7}$, while the following alternating projections brought us to $f(\psi_{N-1}) < 10^{-13}$. These alternating projections changed the prescribed moduli values by less than $10^{-6}$. For example, the true modulus of the fifth sub-diagonal is $1.35 \times 10^{-17}$, which after alternating projections becomes equal to $7.055 \times 10^{-7}$. The details of this example are introduced in TABLE I.

TABLE I.

| True ($T_N$) modulus | Reconstructed modulus | Error |
|---|---|---|
| 0.500000001 | 0.500000001 | 1.75415E-14 |
| 0.356887068 | 0.356886959 | 1.0894E-07 |
| 0.082102303 | 0.082102029 | 2.73629E-07 |
| 0.112727606 | 0.112728148 | 5.42234E-07 |
| 0.087751381 | 0.087752136 | 7.54702E-07 |
| 1.35569E-17 | 7.05501E-07 | 7.05501E-07 |
| 0.035164148 | 0.035164538 | 3.9031E-07 |
| 0.028365603 | 0.028363711 | 1.8926E-06 |
| 0.017520035 | 0.01751814 | 1.89503E-06 |
| 0.023689577 | 0.023689644 | 6.63555E-08 |
| 1.51053E-17 | 7.92723E-07 | 7.92723E-07 |
| 0.034251844 | 0.034250953 | 8.91368E-07 |
| 0.02783223 | 0.02783316 | 9.30331E-07 |
| 0.016923118 | 0.016922442 | 6.76407E-07 |
| 0.027195798 | 0.027196911 | 1.11237E-06 |
| 1.92315E-17 | 1.167E-06 | 1.167E-06 |
| 0.020282772 | 0.020281636 | 1.1363E-06 |
| 0.006067973 | 0.006066215 | 1.75831E-06 |
| 0.018245571 | 0.018246046 | 4.74289E-07 |
| 0.016700478 | 0.016700734 | 2.5643E-07 |

Note that the reconstructed matrix does not coincide with the true matrix (63)-(65). In order to check that the reconstructed matrix belongs to the family (8), we divided the derived matrix element-wise by the elements of the true matrix (63).

$$\mathbf{A} = \left[ \frac{trec_{p-l}}{t_{p-l}} \right] \quad p, l = 1, \ldots, N \quad (66)$$

The matrix $\mathbf{A}$ in (66) is proven to be a rank-one matrix with identical moduli of its eigenvector elements, and phases that demonstrate an ideal linear progression with step-wise $\Delta\varphi = 4.5265^o$ (See TABLE II. ).

TABLE II.

| Eigenvector modulus | Eigenvector phase (deg) | Step-wise eigenvector phase diff. |
|---|---|---|
| 0.223609149 | 0 | 355.473478 |
| 0.223609271 | -4.526521979 | 355.4734728 |
| 0.223609535 | -9.053049216 | 355.475535 |
| 0.223605565 | -13.57751419 | 355.4739444 |
| 0.223604759 | -18.10356975 | 355.4739357 |
| 0.223604759 | -22.62963401 | 355.4732635 |
| 0.223605078 | -27.1563705 | 355.4736034 |
| 0.223605306 | -31.68276709 | 355.4740983 |
| 0.223607431 | -36.20866879 | 355.4740905 |
| 0.223607125 | -40.7345783 | 355.4739358 |
| 0.223607125 | -45.26064254 | 355.4740905 |
| 0.223607431 | -49.78655205 | 355.4740983 |
| 0.223605306 | -54.31245374 | 355.4736034 |
| 0.223605078 | -58.83885033 | 355.4732635 |
| 0.223604759 | -63.36558682 | 355.4739357 |
| 0.223604759 | -67.89165108 | 355.4739444 |
| 0.223605565 | -72.41770664 | 355.475535 |
| 0.223609535 | -76.94217162 | 355.4734728 |
| 0.223609271 | -81.46869886 | 355.473478 |
| 0.223609149 | -85.99522084 | |

This test demonstrates that the reconstructed Toeplitz Hermitian matrix accurately fits the family (8) description.

As discussed in Introduction, along with the successful examples of the Hermitian Toeplitz matrix reconstruction, we came across the unsuccessful reconstructions as well. In TABLE III. we summarize the results of this example for $W_1 = 0.25$, $W_2 = 0.1$ and $\theta_o = 20^o$.

TABLE III.

| True ($T_N$) modulus | Reconstructed modulus | Error |
|---|---|---|
| 0.600000001 | 0.600000001 | 2.58016E-13 |
| 0.372064086 | 0.37182093 | 0.000243157 |
| 0.075682673 | 0.075794101 | 0.000111428 |
| 0.156443831 | 0.15679952 | 0.000355689 |
| 0.023387232 | 0.023484024 | 9.6792E-05 |
| 0.063661977 | 0.063937065 | 0.000275088 |
| 0.015591488 | 0.01563061 | 3.91224E-05 |

| True ($T_N$) modulus | Reconstructed modulus | Error |
|---|---|---|
| 0.056368313 | 0.056746907 | 0.000378594 |
| 0.018920668 | 0.019290562 | 0.000369893 |
| 0.045520357 | 0.045677902 | 0.000157546 |
| 2.0835E-17 | 0.000184868 | 0.000184868 |
| 0.023421279 | 0.023596098 | 0.000174819 |
| 0.012613779 | 0.012 | 0.000613779 |
| 0.028825497 | 0.028403765 | 0.000421732 |
| 0.006682066 | 0.006743835 | 6.17689E-05 |
| 0.021220659 | 0.020785167 | 0.000435492 |
| 0.005846808 | 0.00510211 | 0.000744698 |
| 0.01231112 | 0.013361678 | 0.001050558 |
| 0.008409186 | 0.008503377 | 9.41915E-05 |
| 0.017485663 | 0.017160575 | 0.000325088 |

Our Newton's algorithm converged to only $f\left(\boldsymbol{\psi}_{N-1}^{(t+1)}\right) < 0.0015$, while the following alternating projections resulted in $f\left(\boldsymbol{\psi}_{N-1}^{(t+1)}\right) < 7.47 \times 10^{-12}$. The moduli of the matrix's element are also less accurate ($\Delta \sim 10^{-4}$). The most significant result is that after the Hadamard (element-wise) division, the matrix $\mathbf{A}$ is not a rank-one matrix. The eigenvector that corresponds to the maximal eigenvalue of the matrix $\mathbf{A}$ does not have the same elements moduli, while the inter-element phase difference is not a constant-phase. The less than ideal accuracy of this example may not allow for a categorical analytical judgment, but from a practical viewpoint, the reconstructed matrix is very close to the given matrix, but does not belong to the family (8) of solutions.

Note that the Hermitian Toeplitz matrix reconstruction problem we formulated with ($2N - 1$) given positive values consists of the $N$ moduli of the covariance matrix elements and ($N-1$) eigenvalues. For the given trace of the Hermitian matrix the last eigenvalue is defined by the ($N-1$) preceding ones. But a Hermitian Toeplitz matrix is specified by $2(N-1)$ real-valued parameters and one positive one. Therefore, the revealed existence of the solutions that do not belong to the family (8) should not be too surprising. For the same reason, we may expect different results for the real-valued case, since the numbers of unknown and provided parameters agree in this case.

In this regard, let us be reminded of the H. J. Landau's existence theorem for the real symmetric Toeplitz matrix [9]. According to the theorem 5.1 in [13], "every set of $n$ real numbers is the spectrum of a $n \times n$ real symmetric Toeplitz matrix". Therefore, the eigenvalues alone do not specify the Toeplitz symmetric matrix, and the provided moduli still leave the family of solutions not restricted to (8). Fortunately, the way we checked the reconstructed Hermitian matrix suggests an alternative formulation of the Hermitian Toeplitz matrix reconstruction problem.

If the element-wise division of the given Hermitian matrix and the reconstructed Toeplitz matrix is a rank-one matrix with identical moduli of the eigenvector elements, then the Toeplitz matrix should belong to the family (8). Let $|\mathbf{T}_N|^{-1}$ be the symbol

for the matrix with the inverted moduli of the matrix $\mathbf{R}_N$. Then the matrix $\mathbf{C}_N$

$$\mathbf{C}_N = \boldsymbol{B}_N \otimes |\boldsymbol{T}_N|^{-1} \qquad (67)$$

has all its moduli elements equal to one. Therefore, the element-wise matrix division results may be presented as

$$\mathbf{C}_N = \mathbf{I}_N + \mathbf{R}_N^{\cdot} \otimes \left[\sum_{K=1}^{N-1} \mathbf{A}_K(\boldsymbol{\psi}_K)\right] \qquad (68)$$

where

$$\mathbf{R}_N^{\cdot} = \left[\frac{r_{pl}}{|r_{pl}|}\right], \quad p, l = 1, \dots, N, \qquad (69)$$

and $\mathbf{A}_K(\boldsymbol{\psi}_K)$ is the same as per (43). The symbol $\otimes$ in (68) means the element-wise (Hadamard) matrix product. The equations regarding the eigenvalues of the matrix $\mathbf{C}_N$ that we want to achieve once again involve ($N-1$) eigenvalues of the matrix $\mathbf{C}_N$ with a constant main diagonal:

$$f_{N-1}(\boldsymbol{\psi}_{N-1}) \coloneqq \begin{bmatrix} \lambda_2(\boldsymbol{\psi}_{N-1}) \\ \vdots \\ \lambda_N(\boldsymbol{\psi}_{N-1}) \end{bmatrix} = 0 . \qquad (70)$$

If (70) is true, then the first eigenvalue of the matrix $\mathbf{C}$ is equal to

$$\lambda_1(\boldsymbol{\psi}_{N-1}) = \mathrm{Tr}\, \mathbf{C}_N = N \qquad (71)$$

Similar to (51), the Newton's algorithm for the solution in (70) is:

$$\mathbf{J}(\boldsymbol{\psi}_{N-1}^{(t)})\left[\boldsymbol{\psi}_{N-1}^{(t+1)} - \boldsymbol{\psi}_{N-1}^{(t)}\right] = -f_{N-1}(\boldsymbol{\psi}_{N-1}^{(t)}) \qquad (72)$$

$$\left[\mathbf{J}(\boldsymbol{\psi}_N^{(t)})\right]_{jk} = \frac{\partial \lambda_j\left[\mathbf{C}_N(\boldsymbol{\psi}_{N-1}^{(t)})\right]}{\partial \psi_k} \qquad (73)$$
$$j = 2, \dots, N$$
$$k = 1, \dots, N-1$$

$$\mathbf{J}_{jk} = \mathbf{U}_j^{\mathrm{H}}(\boldsymbol{\psi}_{N-1})\left[\mathbf{R}_N^{\cdot} \otimes \mathbf{B}_k(\boldsymbol{\psi}_k)\right]\mathbf{U}_j(\boldsymbol{\psi}_{N-1}) \qquad (74)$$

For the invertible square Jacobian (73), (74), we get:

$$\boldsymbol{\psi}_{N-1}^{(t+1)} = \boldsymbol{\psi}_{N-1}^{(t)} - \mathbf{J}_{N-1}^{-1}(\boldsymbol{\psi}_{N-1}^{(t)})f_{N-1}(\boldsymbol{\psi}_{N-1}^{(t)}) \qquad (75)$$

Note that the same control over the phase innovation

$$\max_k \left| \mathbf{J}_{N-1}^{-1}(\boldsymbol{\psi}_{N-1}^{(t)})f_{N-1}(\boldsymbol{\psi}_{N-1}^{(t)}) \right| \leq 15^o \qquad (76)$$

should be considered to retain the affine approximation in every iteration.

Obviously, if we achieve the solution of the equation (70), the reconstructed $N$-variate Toeplitz matrix should belong to the family (8). In fact, in all conducted trials reported below, we got the desired results. Computations were conducted for the matrix (63), for $W_1 = 0.2$, $W_2 = 0.1$ and for $W_1 = 0.25$, $W_2 = 0.1$, including the case with the reported failure reported to get the family (8) solution. For each trial, we used the independent phase errors $\varphi_n$ and the phase offset $\theta_o$. The results are presented in TABLE IV. Here we introduce the moduli of the "non-zero" eigenvector of the matrix $\mathbf{C}_N$, its phases, and the inter-element phase difference after subtraction of the random phase errors $\varphi_n$, $n = 1, \dots, N-1$.

TABLE IV.

| Eigenvector modulus | Eigenvector phase (deg) | Step-wise eigenvector phase diff. |
|---|---|---|
| 0.223606798 | 0 | 60.23741136 |
| 0.223606798 | 60.23741136 | 60.23741136 |
| 0.223606798 | 120.4748227 | 60.23741136 |
| 0.223606798 | 180.7122341 | 60.23741136 |
| 0.223606798 | -119.0503545 | 60.23741136 |
| 0.223606798 | -58.81294318 | 60.23741136 |
| 0.223606798 | 1.424468183 | 60.23741136 |
| 0.223606798 | 61.66187955 | 60.23741136 |
| 0.223606798 | 121.8992909 | 60.23741136 |
| 0.223606798 | -177.8632977 | 60.23741136 |
| 0.223606798 | -117.6258864 | 60.23741136 |
| 0.223606798 | -57.388475 | 60.23741136 |
| 0.223606798 | 2.848936366 | 60.23741136 |
| 0.223606798 | 63.08634773 | 60.23741136 |
| 0.223606798 | 123.3237591 | 60.23741136 |
| 0.223606798 | -176.4388295 | 60.23741136 |
| 0.223606798 | -116.2014182 | 60.23741136 |
| 0.223606798 | -55.96400681 | 60.23741136 |
| 0.223606798 | 4.273404549 | 60.23741136 |
| 0.223606798 | 64.51081591 | |

In all the trials we got a constant modulus eigenvector, with the accurately linear phase progression after compensating for the random-like phase "calibration" errors.

## IV. RECONSTRUCTION OF THE REAL-VALUED TOEPLITZ SYMMETRIC MATRIX, GIVEN THE MODULI OF ITS ELEMENTS AND THE SET OF N EIGENVALUES

In this reconstruction problem we are given the moduli of the symmetric Toeplitz matrix that form the initial positive-valued Toeplitz matrix $\mathbf{T}_N^{(o)}$:

$$\mathbf{T}_N^{(o)} = \text{Toep}\,[t_o, |t_1|, \dots, |t_{N-1}|]\,, \qquad (77)$$

where Toep $[x_0, x_1, \dots, x_{N-1}]$ is the operation that forms the $N$-variate Hermitian Toeplitz matrix, given a set of one positive number $x_0$ and $(N-1)$ complex numbers $x_j, j = 1, \dots, N-1$. For $x_j \geq 0$, this operator creates a symmetric real-valued Toeplitz matrix with

$$t_{p-l} = t_{l-p}\,(t_{p-l} = k) \qquad (78)$$

As we mentioned in Introduction, we have to distribute an unspecified number of sign inversions to the elements $|t_1|, \dots, |t_{N-1}|$, so that the Toeplitz symmetric matrix specified by this modified sequence has the prescribed set of eigenvalues. First, note that not every positive-valued Toeplitz symmetric matrix $\mathbf{T}_N^{(o)}$ may be transformed into a Toeplitz symmetric matrix with the prescribed eigenvalues.

As mentioned above, a single example provided by Prof. M. T. Chu [33] is sufficient to demonstrate this statement. Let us consider the symmetric (non-Toeplitz!) matrix $\mathbf{A}$:

$$\mathbf{A} = \begin{bmatrix} 10 & 1 & 1 & 1 & -1 & -1 \\ 1 & 10 & -1 & -1 & -1 & 1 \\ 1 & -1 & 10 & 1 & -1 & 1 \\ 1 & -1 & -1 & 10 & -1 & -1 \\ -1 & -1 & -1 & -1 & 10 & -1 \\ -1 & 1 & 1 & -1 & -1 & 10 \end{bmatrix} \qquad (79)$$

This matrix has eigenvalues

$$7.000\,;\,7.639\,;\,9.000\,;\,11.000\,;\,12.2361\,;\,13.000 \qquad (80)$$

The matrix $\mathbf{A}^{(o)}$, constructed of the elements moduli, is obviously a Toeplitz symmetric matrix represented as

$$\mathbf{A}^{(o)} = 9\mathbf{I}_o + \mathbf{e}_6\mathbf{e}_6^{\mathsf{T}} \qquad (81)$$

$\mathbf{e}_6^{\mathsf{T}} = (1, 1, \dots, 1)$. This matrix has five eigenvalues equal to nine and one eigenvalue equal to 15. The reconstruction technique described below resulted in the Toeplitz matrix $\mathbf{A}_{TOEP}$:

$$\mathbf{A}_{TOEP} = \text{Toep}\,[10, -1, 1, 1, -1, 1] \qquad (82)$$

with eigenvalues

$$7.3961\,;\,7.77639\,;\,9.000\,;\,10.1099\,;\,12.2361\,;\,13.1940 \qquad (83)$$

which are close to (80) but still very different. This example demonstrates that for the method to work, the provided eigenvalues should indeed belong to a Hermitian matrix related to a Toeplitz real-valued symmetric matrix by the transformation (8) that destroys the Toeplitz structure, but retains the eigenvalues.

As discussed in Introduction, the most important thing is the question about the family of these ToIEP problem solutions. In other words, the question is how many real-valued symmetric Toeplitz matrices exist with the same eigenvalues and elements' moduli. This question was addressed in [4] by the analysis of the stability group of symmetric Toeplitz matrices. Let the symmetric Toeplitz matrix $\mathbf{T}(\mathbf{r})$ be specified by the entries of its first column $\mathbf{r} \in \mathbb{R}^N$. The linear subspace $\mathbf{J}(N)$ of all symmetric Toeplitz matrices is spanned by the matrices

$$\mathbf{E}^{(i)} \coloneqq \mathbf{T}(l^{(i)}) \tag{84}$$

where $l^{(i)}, i = 1, \dots, N$ is the $i$-th standard basis in $\mathbb{R}^N$. With respect to the Frobenius norm, the elements in the basis $\{\mathbf{E}^{(i)}\}$ are mutually orthogonal. The stability group $\mathbf{S}(N)$ of $\mathbf{J}(N)$ is defined in [4] to be:

$$\begin{aligned}\mathbf{S}(N) \\ \coloneqq \{\mathbf{Q} \in \mathbf{O}(N) | \mathbf{Q}(N)\mathbf{T}\mathbf{Q}^{\mathrm{T}}(N) \in \mathbf{J}(N), \text{ if } \mathbf{T} \in \mathbf{J}(N)\}\end{aligned} \tag{85}$$

The name "group" was justified in [4] by the fact that for each $\mathbf{Q} \in \mathbf{S}(N)$, the action $\mathbf{Q}\mathbf{T}\mathbf{Q}^{\mathrm{T}}$ is the bijection from $\mathbf{J}(N)$ to itself. Obviously, the identity matrix $\mathbf{I}_N$ and the backward identity matrix $\mathbf{J}_N$ ,

$$\mathbf{J}_N = \begin{bmatrix} 0 & & & 1 \\ & & \ddots & \\ & 1 & & \\ 1 & & & 0 \end{bmatrix}, \tag{86}$$

are in $\mathbf{S}(N)$. Yet, in the Theorem 2.3 in [4] M. T. Chu proved the following theorem.

Theorem 2.3.    Suppose $N > 1$. The stability group $\mathbf{S}(N)$ of $\mathbf{J}(N)$ has exactly eight elements, regardless of the dimension $N$. The elements are $\pm \mathbf{I}_N, \pm \mathbf{J}_N, \pm \mathbf{I}'_N, \pm \mathbf{J}'_N$, where

$$\mathbf{I}'_N = \begin{bmatrix} -1 & & & & 0 \\ & +1 & & & \\ & & -1 & & \\ & & & \ddots & \\ & & & & (-1)^{N-1} & \\ 0 & & & & & (-1)^N \end{bmatrix};$$

$$\begin{bmatrix} 0 & & & & & -1 \\ & & & & -1 & +1 \\ & & & \ddots & & \\ & & (-1)^{N-1} & & & \\ & \ddots & & & & \\ (-1)^N & & & & & 0 \end{bmatrix} = \mathbf{J}'_N \tag{87}$$

Since the backward identity matrix reverses the order of the antenna array elements, it is not an option for the array calibration problem. Therefore we may safely conclude that apart from the given symmetric Toeplitz matrix $\mathbf{T}_N$

$$\mathbf{T}_N = (\pm \mathbf{I}_N)\mathbf{T}_N(\pm \mathbf{I}_N), \tag{88}$$

only the transformation

$$\mathbf{I}'_N \mathbf{T}_N \mathbf{I}'_N \tag{89}$$

may have the same eigenvalues and elements moduli as the matrix $\mathbf{T}_N$. Therefore, as the result of the symmetric Toeplitz matrix reconstruction, we can get one of the two isomorphic solutions (88) or (89). Note that

$$\mathbf{I}'_N = \mathrm{diag}\,[\exp(j\pi n)]\;,\; n = 1, \dots, N \tag{90}$$

which means that for $d/\lambda \geq 1/2$, the angular spectrum symmetric with respect to $\theta_o = 0^\mathrm{o}$, got shifted by $\Delta\theta$:

$$\frac{d}{\lambda}\sin\Delta\theta = \frac{1}{2} \tag{91}$$

with $\Delta\theta = 90^\mathrm{o}$ for $d/\lambda = 1/2$. This property may be used to discriminate between the two isomorphic solutions.

Let us introduce our optimization algorithm starting from the formal optimization criterion:

$$\min\max_n|\lambda_n(k) - \lambda_n^*| = \Delta(k) \tag{92}$$

where $\lambda_n^*$, $n = 1,\dots, N$ are the provided eigenvalues, and $\lambda_n(k), n = 1,\dots, N$ are the eigenvalues of the reconstructed Toeplitz matrix at the $k$-th step of the reconstruction. For the integer optimization problem of assigning the sign changes to the matrix $\mathbf{T}_N^{(o)}$ sub-diagonals, we use the simplest "maximum element" technique (see [34], for example). Let the $(N-1)$ variate vector of the non-negative elements $\mathbf{X}_{N-1}(0) \geq 0$ represent the first row of the initial Toeplitz matrix $\mathbf{T}_N^{(o)}$ compiled with the moduli of the matrix's $\mathbf{R}_N[f(\theta)]$ elements. The diagonal element $x_o$ remains positive and unmodified.

Let $l_{jk}$ be a symbol of the sign change in the $j$-th element of the vector $\mathbf{X}_{N-1}(k-1)$ that forms the Toeplitz symmetric matrix on the previous $(k-1)$-st step of optimization. Let $\Lambda[l_j, k]$ be the criterion (92) value achieved due to the sign change in the $j$-th element of the vector $\mathbf{X}_{N-1}(k-1)$ at the $k$-th step. Then, the "maximum element" algorithm may be formalized as follows:

$$\begin{aligned}\text{Find } \min \Lambda[l_j, k] \\ j = 1, \dots, N-1 \\ k = 1, 2, \dots\end{aligned} \tag{93}$$

The algorithm (93) is quite simple. At the $k$-th step we find the best (out of $N-1$ possible) element in the vector $\mathbf{X}_{N-1}(k-1)$ with the sign inversion in the sub-diagonal associated with this element that leads to the best gain in the optimization criterion. Iteration continues until the optimum solution with $\Lambda[\mathrm{k}] = 0$ is found, or no further criterion improvement can be achieved. This method has a number of modifications, including the "dynamic programming" technique first applied in [20]. Within this method, for each $(N-1)$ initial sign inversions in the vector $\mathbf{X}_{N-1}(0)$, the algorithm described above is applied with potentially up to $(N-1)$ different solutions, from which the best one can be chosen.

While little is proven about the computational efficiency of the "maximum element" algorithm, its application for the Toeplitz matrix reconstruction is warranted by two considerations. First is that the optimum solution with $\Lambda[k] =$ 0 is known. Second is that no constraints on the number of admissible sign changes exist. In the next Sec. V we provide the results of the numerical trials with the accurately known matrices, where the "maximum element" algorithm converged to the optimum solution $\Lambda[k] = 0$ over a very modest number of steps.

Finally, note that the same algorithm may be applied to the MRA antenna arrays with the real-valued covariance matrices. The only difference is that in the MRA array case, the optimization criterion checks for convergence of the $M < N$ eigenvalues of the matrix (35):

$$\mathbf{R}_M = \mathbf{H}_{M,N}\mathbf{T}_N[f(\theta)]\mathbf{H}_{M,N}^T \qquad (94)$$

As mentioned above, despite the "fully augmentable" property of the considered MRA arrays, reductions of the provided data from $(2N-1)$ positive values for the full ULA array case to $(N+M)$ positive numbers in the MRA case may not allow for the true solution for the $N$-variate Toeplitz matrix to be found. In the special case, when the signal subspace dimension of the matrix $\mathbf{R}_M$ is less than $M$, so that

$$\lambda_{min}[\mathbf{R}_M] = \sigma_{\omega n}^2 \qquad (95)$$

where $\sigma_{\omega n}^2$ is the additive white noise power, the $M$ eigenvalues of the matrix $\mathbf{R}_M$ may be augmented by the $(N-M)$ equal to $\sigma_{\omega n}^2$ eigenvalues of the $N$-variate matrix $\mathbf{T}_N$. With the matching number of provided parameters and unknowns, we may expect to receive the same (two) solutions to the Toeplitz restoration problem. Numerical results are provided in the next Section V.

## V. NUMERICAL RESULTS FOR THE REAL-VALUED SYMMETRIC TOEPLITZ MATRIX RECONSTRUCTION

Below we provide the numerical analysis of the Toeplitz symmetric matrix reconstruction for the two matrices specified in (24) and (29). We start our analysis from the sinc($W$) matrix (24) reconstruction for $W = 0.2$, $0.25$, $0.3$, $0.35$ and $0.4$, covering the broad range of the rectangular spectrum beamwidth values. For $N = 20$, the numerical results are illustrated by Fig. 1-Fig. 5 for each $W$ correspondingly. Each figure has two sub-figures: the top figure demonstrates the element-wise mismatch between the elements of the true matrix, sinc($W$), and the two isomorphic solutions, $\mathbf{T}_N$ and $\mathbf{D}(\pi)\mathbf{T}_n\mathbf{D}(\pi)$; the bottom figure illustrates the original eigenspectrum of the sinc($W$) matrix and the eigenspectrum of the reconstructed matrix.

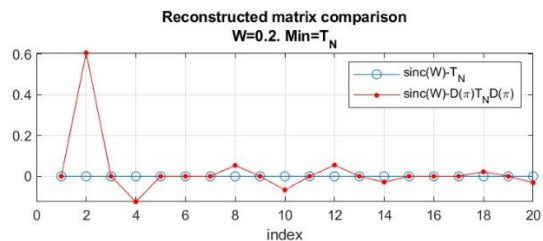

Fig. 1. Real-valued sinc($W$) vs. reconstruction. N=20. W=0.2.

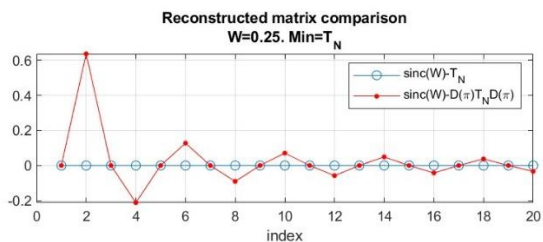

Fig. 2. Real-valued sinc($W$) vs. reconstruction. N=20. W=0.25.

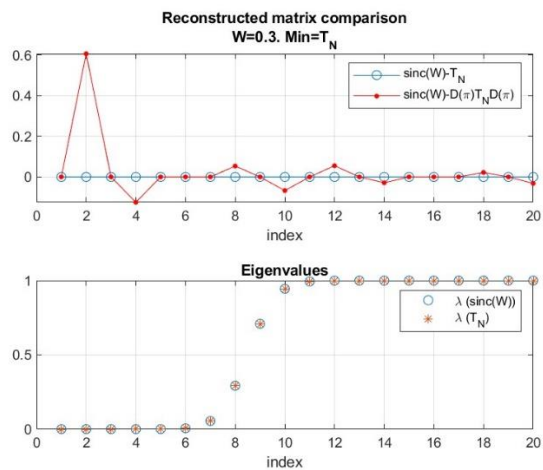

Fig. 3. Real-valued sinc($W$) vs. reconstruction. N=20. W=0.3.

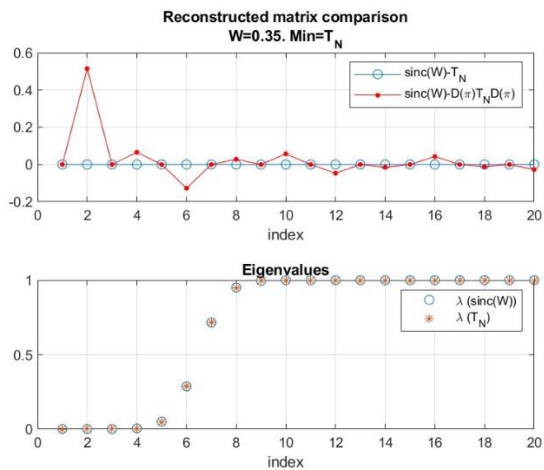

Fig. 4. Real-valued sinc($W$) vs. reconstruction. N=20. W=0.35.

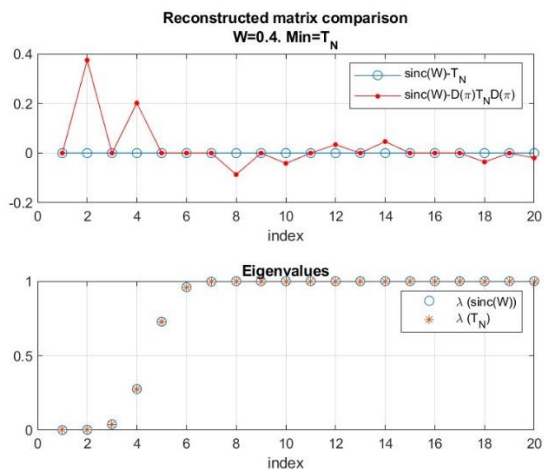

Fig. 5. Real-valued sinc($W$) vs. reconstruction. N=20. W=0.4.

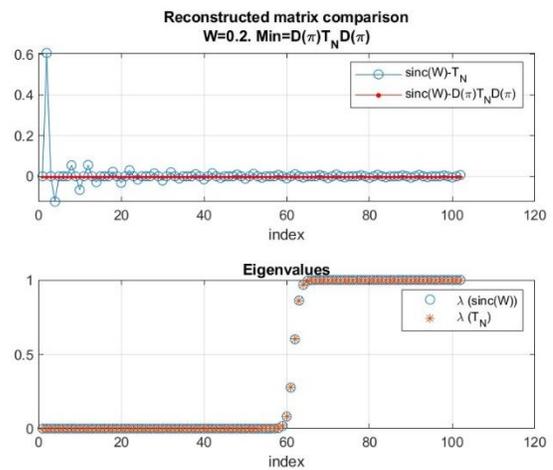

Fig. 6. Real-valued sinc($W$) vs. reconstruction. N=102. W=0.2.

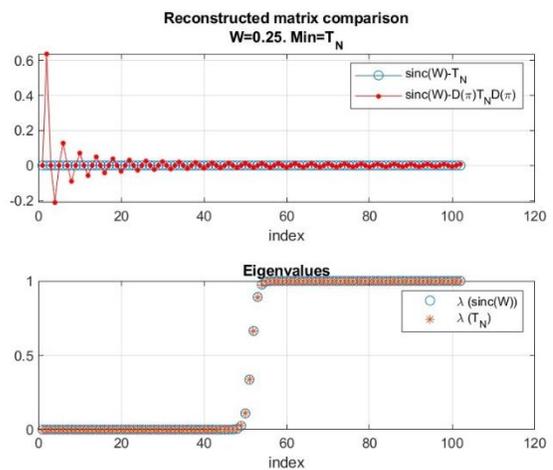

Fig. 7. Real-valued sinc($W$) vs. reconstruction. N=102. W=0.25.

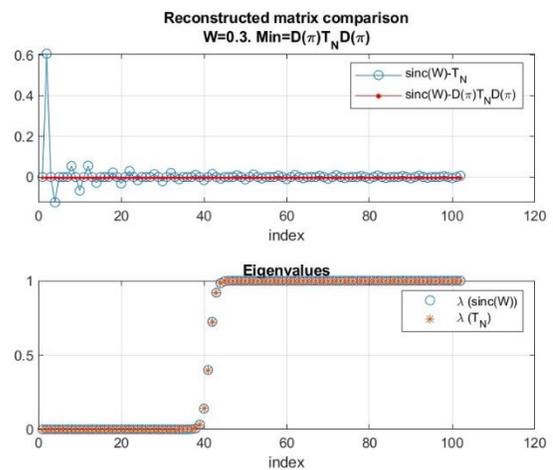

Fig. 8. Real-valued sinc($W$) vs. reconstruction. N=102. W=0.3.

One can see that in all considered cases, the "maximum element" optimization algorithm led to the ideal match of the eigenvalues of the original and the reconstructed Toeplitz matrix.

The actual element values of the original sinc($W$) matrix ideally coincided with the reconstructed Toeplitz matrix elements in all cases with $W$ = 0.2 - 0.4 for $N$ = 102. Fig. 6-Fig. 10 in the same format illustrates these numerical results. As one can see, for $N$ = 102 we once again got the ideal matching eigenvalues of the original and reconstructed Toeplitz matrices. In all cases but $W$ = 0.3, we got the ideal match of the reconstructed matrix $\mathbf{T}_N$ with the sinc($W$) matrix. For the case with $W$ = 0.3, the ideal match between the elements of the original sinc($W$) matrix was observed with the elements of the isomorphic solution $\mathbf{D}(\pi)\mathbf{T}_n\mathbf{D}(\pi)$. Therefore, the possibility to reconstruct one of the two isomorphic solutions has been experimentally validated.

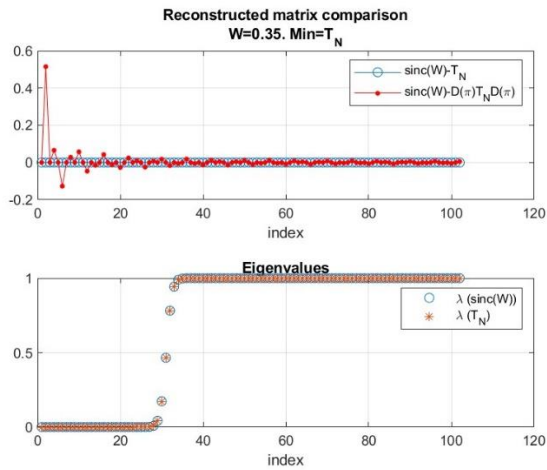

Fig. 9. Real-valued sinc($W$) vs. reconstruction. N=102. W=0.35.

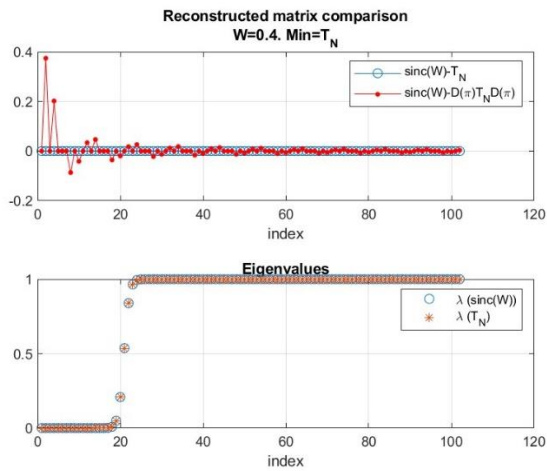

Fig. 10. Real-valued sinc($W$) vs. reconstruction. N=102. W=0.4.

Next, in Fig. 11-Fig. 30 we report the numerical results for the matrix (29), for the same set of $W$ values ($W$ = 0.2 - 0.4) and $\alpha$ = 0.1, 1, 5, 10 for $N$ = 20.

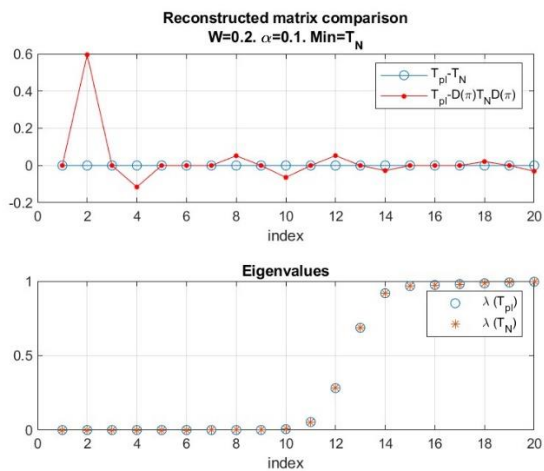

Fig. 11. $t_{p\text{-}1}$ vs. reconstruction. N=20. W=0.2. $\alpha$=0.1.

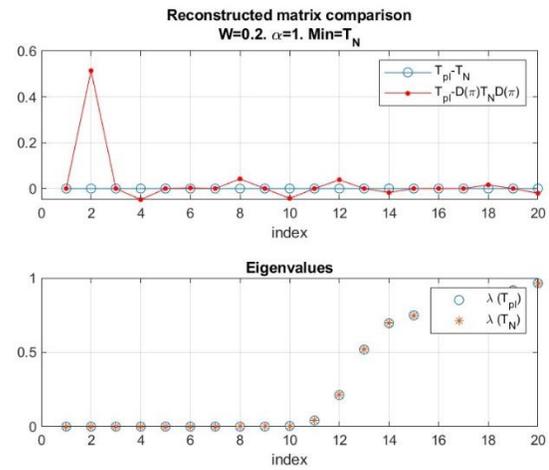

Fig. 12. $t_{p\text{-}1}$ vs. reconstruction. N=20. W=0.2. $\alpha$=1.

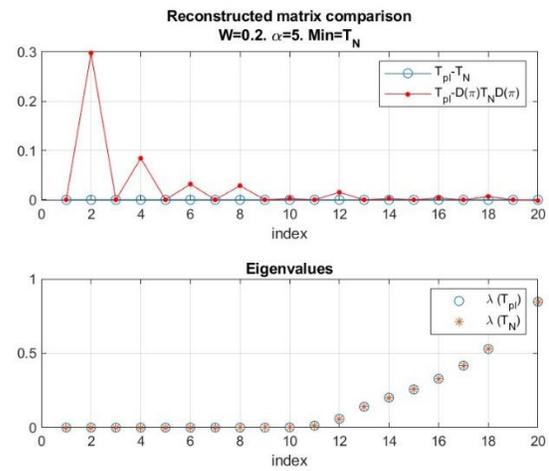

Fig. 13. $t_{p\text{-}1}$ vs. reconstruction. N=20. W=0.2. $\alpha$=5.

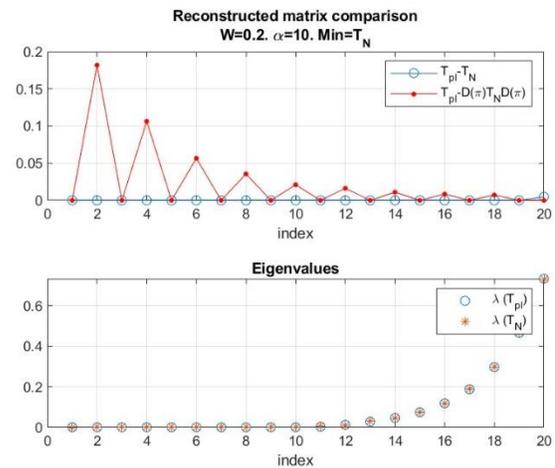

Fig. 14. $t_{p\text{-}1}$ vs. reconstruction. N=20. W=0.2. $\alpha$=10.

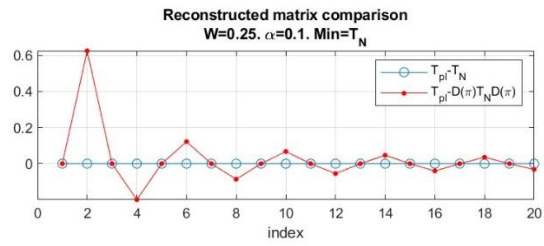

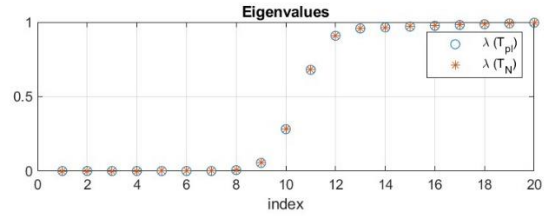

Fig. 15. $t_{p-l}$ vs. reconstruction. N=20. W=0.25. $\alpha$=0.1.

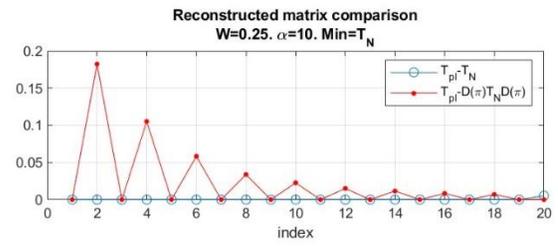

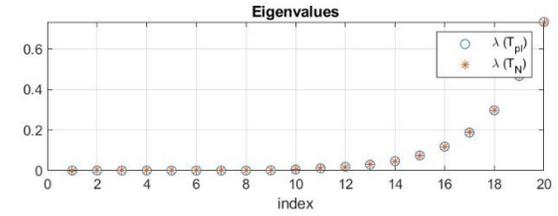

Fig. 18. $t_{p-l}$ vs. reconstruction. N=20. W=0.25. $\alpha$=10.

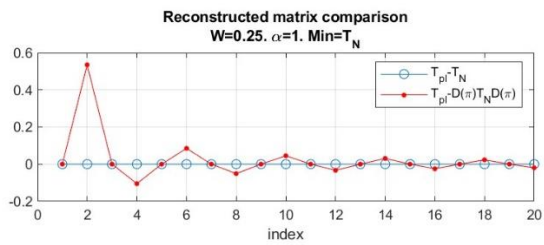

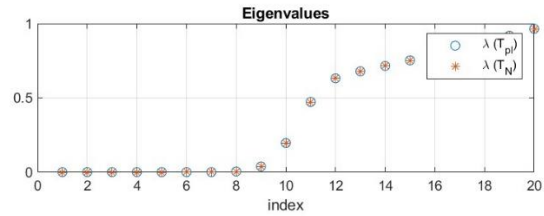

Fig. 16. $t_{p-l}$ vs. reconstruction. N=20. W=0.25. $\alpha$=1.

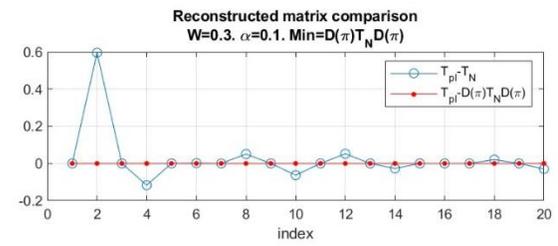

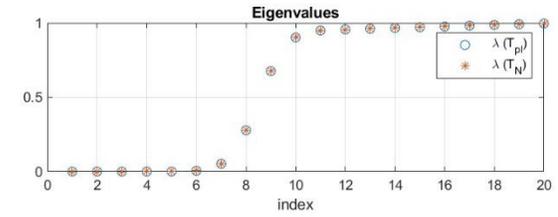

Fig. 19. $t_{p-l}$ vs. reconstruction. N=20. W=0.3. $\alpha$=0.1.

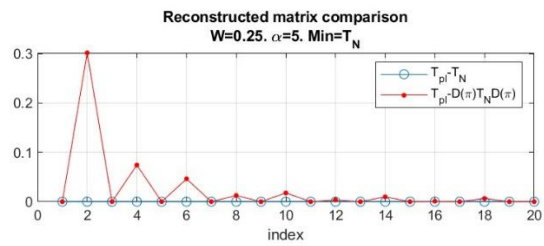

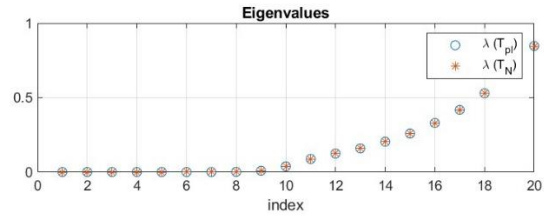

Fig. 17. $t_{p-l}$ vs. reconstruction. N=20. W=0.25. $\alpha$=5.

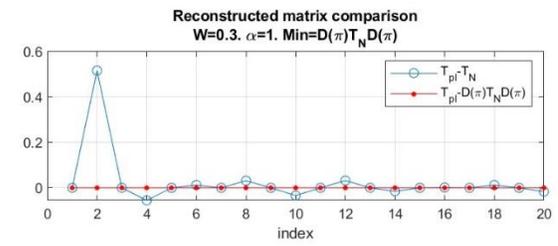

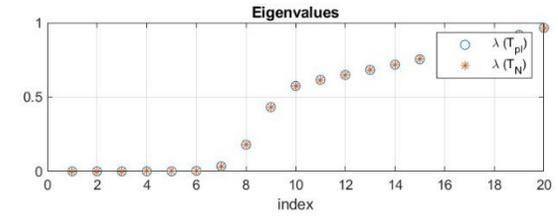

Fig. 20. $t_{p-l}$ vs. reconstruction. N=20. W=0.3. $\alpha$=1.

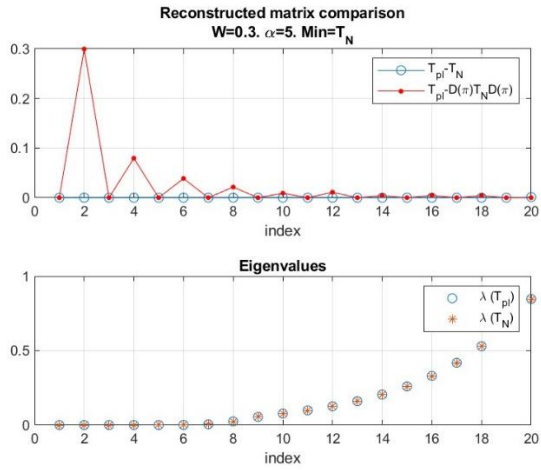

Fig. 21. $t_{p\text{-}l}$ vs. reconstruction. N=20. W=0.3. α=5.

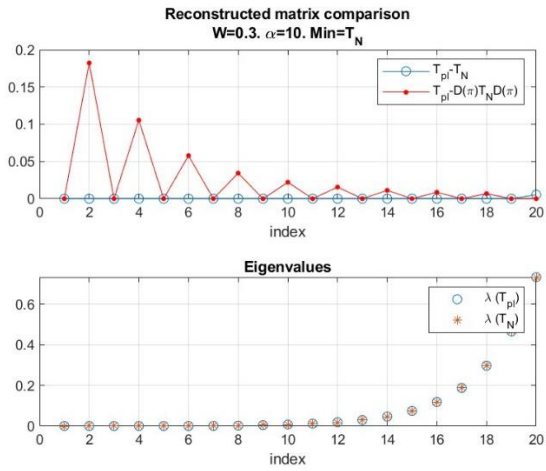

Fig. 22. $t_{p\text{-}l}$ vs. reconstruction. N=20. W=0.3. α=10.

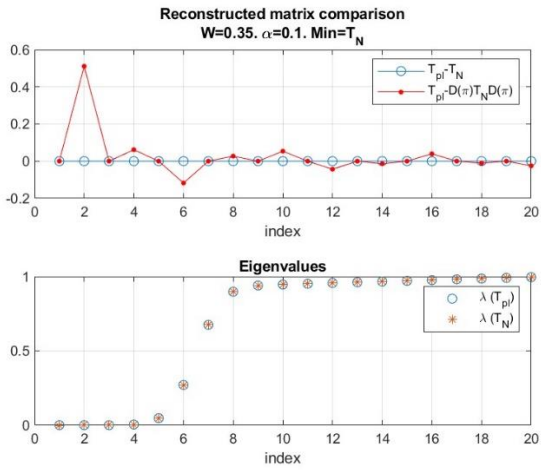

Fig. 23. $t_{p\text{-}l}$ vs. reconstruction. N=20. W=0.35. α=0.1.

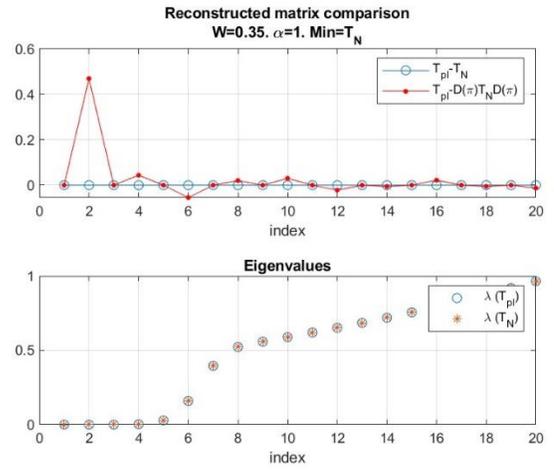

Fig. 24. $t_{p\text{-}l}$ vs. reconstruction. N=20. W=0.35. α=1.

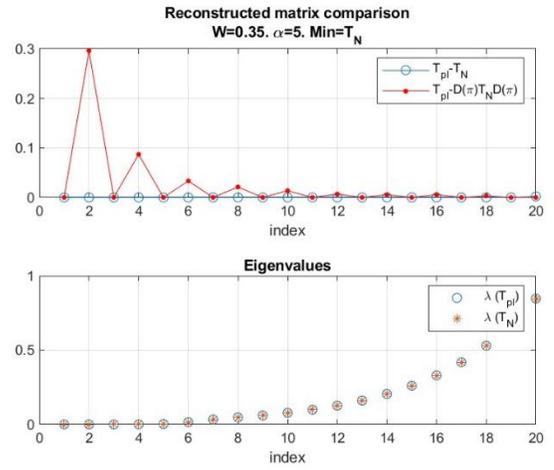

Fig. 25. $t_{p\text{-}l}$ vs. reconstruction. N=20. W=0.35. α=5.

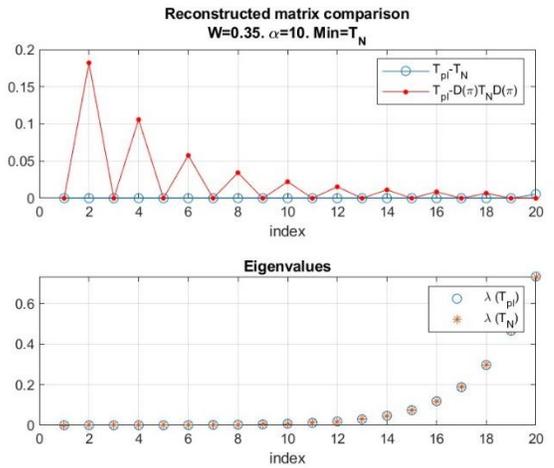

Fig. 26. $t_{p\text{-}l}$ vs. reconstruction. N=20. W=0.35. α=10.

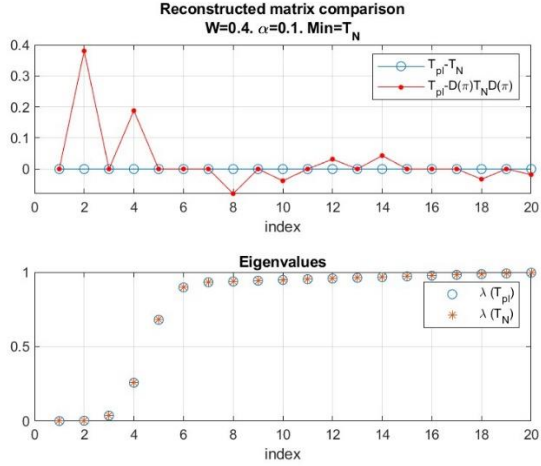

Fig. 27. $t_{p\text{-}l}$ vs. reconstruction. N=20. W=0.4. $\alpha$=0.1.

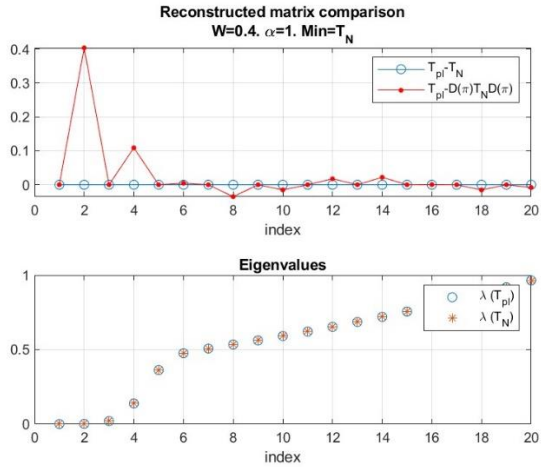

Fig. 28. $t_{p\text{-}l}$ vs. reconstruction. N=20. W=0.4. $\alpha$=1.

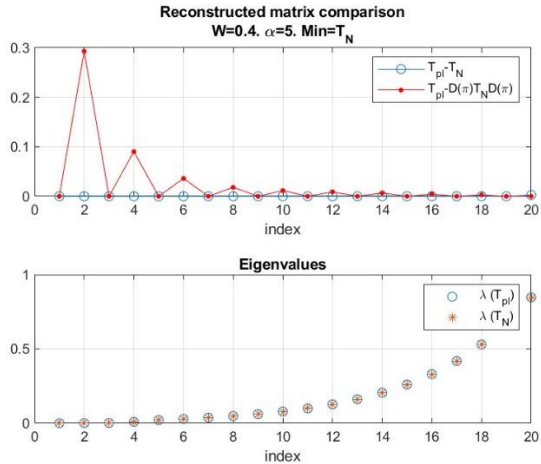

Fig. 29. $t_{p\text{-}l}$ vs. reconstruction. N=20. W=0.4. $\alpha$=5.

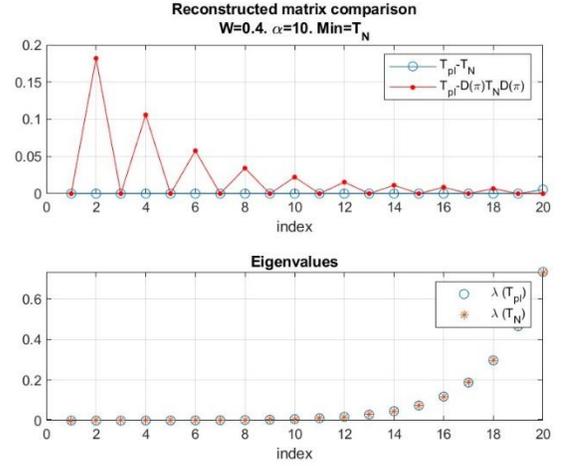

Fig. 30. $t_{p\text{-}l}$ vs. reconstruction. N=20. W=0.4. $\alpha$=10.

For only two cases ($W = 0.3$, $\alpha = 0.1$ and $\alpha = 1$), the ideal match was observed between the original (ideal) Toeplitz matrix and the isomorphic solution $\mathbf{D}(\pi)\mathbf{T}_n\mathbf{D}(\pi)$. In all other cases, we observed the ideal match between the given and the reconstructed Toeplitz matrices.

Note that the number of the distributed sign inversions that led to the optimum solution did not exceed the number of negative entries in the first row of the true Toeplitz matrix, which means a close to the optimum convergence rate. Therefore, the provided numerical examples not only validated the theoretical prediction on the existence of the two isomorphic solutions $\mathbf{T}_N$ and $\mathbf{D}(\pi)\mathbf{T}_N\mathbf{D}(\pi)$, but also demonstrated the high computational efficiency of the proposed optimization technique.

Let us now apply the same optimization algorithm for the $N$-element Toeplitz symmetric matrix reconstruction given the $M \ll N$ eigenvalues of the $M$-variate MRA covariance matrix $\mathbf{R}_M[f(\theta)]$ (35). Specifically, let us consider the $M = 17$-element MRA defined by the number [35]:

$$1^2, 3, 5^2, 11^6, 6^3, 1^3 \qquad (96)$$

According to this ruler, the following MRA spans the aperture of the $N = 102$-element ULA:

$$\{0, 1, 2, 5, 10, 15, 26, 37, 48, 59, 70, 81, 87, 93, 99, 100, 101\} \qquad (97)$$

The numbers in (97) specify the positions of "1"s in each of the 17 rows of the (17 x 102)-variate matrix $\mathbf{H}_{17,102}$. Correspondingly, the eigenvalues of the matrix

$$\mathbf{R}_{17} = \mathbf{H}_{17,102}\mathbf{T}_{102}[f(\theta)]\mathbf{H}_{17,102}^{\mathsf{T}} \qquad (98)$$

are provided, while the optimization should be searching for the solution of the equation

$$\begin{aligned} &\text{eig}_m\big[\mathbf{H}_{17,102}\mathbf{T}_{102}(k)\mathbf{H}_{17,102}^{\mathsf{T}}\big] \\ &\quad - \text{eig}_m\big[\mathbf{H}_{17,102}\mathbf{T}_{102}[f(\theta)]\mathbf{H}_{17,102}^{\mathsf{T}}\big] = 0 \end{aligned} \qquad (99)$$

Along with checking convergence of the eigenvalues (99), we also controlled convergence of the reconstructed $N$-variate

Toeplitz matrix to the original matrix $\mathbf{T}_{102}[f(\theta)]$. Results of the numerical trials are illustrated in TABLE V.

TABLE V.

| W=0.2 | | W=0.25 | |
|---|---|---|---|
| $\lambda_{R17}$ | $|\lambda_{R17} - \lambda_{Rrec}|$ | $\lambda_{R17}$ | $|\lambda_{R17} - \lambda_{Rrec}|$ |
| 0.895870107 | 0.001262692 | 0.954533586 | 3.8645E-05 |
| 0.880026666 | 0.00261553 | 0.951616287 | 3.00178E-05 |
| 0.492386434 | 0.000704632 | 0.638020093 | 2.31205E-05 |
| 0.461028513 | 0.002147799 | 0.578338465 | 2.21303E-05 |
| 0.454897668 | 2.84931E-06 | 0.550539424 | 7.20646E-06 |
| 0.435257361 | 0.007705813 | 0.545414008 | 1.12626E-06 |
| 0.403085002 | 0.00330055 | 0.536841847 | 0.000157254 |
| 0.400037368 | 0.001941418 | 0.5 | 1.11022E-16 |
| 0.366997664 | 0.011391232 | 0.5 | 0 |
| 0.363641959 | 0.001324917 | 0.5 | 1.11022E-16 |
| 0.36271115 | 0.000809725 | 0.463158153 | 0.000157254 |
| 0.357085123 | 0.002192214 | 0.454585992 | 1.12626E-06 |
| 0.335121595 | 0.001459982 | 0.449460576 | 7.20646E-06 |
| 0.283709639 | 0.000714949 | 0.421661535 | 2.21303E-05 |
| 0.279289008 | 0.000303546 | 0.361979907 | 2.31205E-05 |
| 0.015723427 | 8.13298E-06 | 0.048383713 | 3.00178E-05 |
| 0.013131315 | 0.002661454 | 0.045466414 | 3.8645E-05 |

| W=0.3 | | W=0.35 | |
|---|---|---|---|
| $\lambda_{R17}$ | $|\lambda_{R17} - \lambda_{Rrec}|$ | $\lambda_{R17}$ | $|\lambda_{R17} - \lambda_{Rrec}|$ |
| 0.986868685 | 0.001018099 | 0.99832086 | 0.00065057 |
| 0.984276573 | 0.00307673 | 0.996550946 | 0.000700554 |
| 0.720710992 | 0.001118598 | 0.891367291 | 0.000320413 |
| 0.716290361 | 0.000500894 | 0.877329819 | 0.000331201 |
| 0.664878405 | 7.6929E-05 | 0.78000808 | 0.000273156 |
| 0.642914877 | 0.000638786 | 0.750383427 | 0.000426316 |
| 0.63728885 | 0.000683095 | 0.727272577 | 0.000655807 |
| 0.636358041 | 0.000259135 | 0.727239279 | 0.000323051 |
| 0.633002336 | 0.001890503 | 0.726090438 | 0.00070729 |
| 0.599962632 | 0.000214902 | 0.699964366 | 0.00041987 |
| 0.596914998 | 0.000223198 | 0.669091814 | 0.000223539 |
| 0.564742639 | 0.000108016 | 0.666498022 | 0.000263651 |
| 0.545102332 | 4.13649E-05 | 0.650475494 | 0.000654066 |
| 0.538971487 | 0.001096339 | 0.636993236 | 0.000175385 |
| 0.507613566 | 0.001283814 | 0.610077675 | 0.001501782 |
| 0.119973334 | 4.47966E-05 | 0.252853877 | 0.000167684 |
| 0.104129893 | 0.000681874 | 0.239482798 | 0.0009865 |

| W=0.4 | |
|---|---|
| $\lambda_{R17}$ | $|\lambda_{R17} - \lambda_{Rrec}|$ |
| 0.999966633 | 0.003173259 |
| 0.999682857 | 0.0007518 |
| 0.987061258 | 0.000883191 |
| 0.953586128 | 0.003472783 |
| 0.833352357 | 0.000184306 |
| 0.829521525 | 0.000772515 |
| 0.823460767 | 0.001675873 |
| 0.818182206 | 0.000143671 |
| 0.818180173 | 0.000461566 |
| 0.817980937 | 0.001429328 |
| 0.80829145 | 0.003134876 |
| 0.799857398 | 0.00010538 |
| 0.794589051 | 0.002307522 |
| 0.741403167 | 0.000739236 |
| 0.719641759 | 0.002043735 |
| 0.433724674 | 0.002597534 |
| 0.42151766 | 0.001750829 |

In TABLE V. , the eigenvalues of the ideal matrix (98) are compared with the eigenvalues of the reconstructed matrix $\mathbf{H}_{17,102}\mathbf{T}_{102}(k)\mathbf{H}_{17,102}^{\mathrm{T}}$. One can see that the eigenvalues of the (17 x 17)-variate MRA covariance (true) matrix are restored with high enough accuracy for all tests of $W = 0.2 - 0.4$. Unfortunately, the demonstrated equality of the eigenvalues is the only property of the true MRA covariance matrix we were able to restore. Neither the (17 x 17)-variate MRA matrix $\mathbf{R}_{17}$, nor the (102 x 102)-variate real-valued Toeplitz matrix (or its isomorphic variant) were restored. The reconstructed MRA matrices with the given matrices' moduli as per the true MRA matrix and the very accurately restored eigenvalues, differ by 0.3742 and 0.20182 (at most) from the true matrix $\mathbf{H}_{17,102}\mathbf{T}_{102}[f(\theta)]\mathbf{H}_{17,102}^{\mathrm{T}}$ and from its isomorphic variant $\mathbf{H}_{17,102}\mathbf{D}(\pi)\mathbf{T}_{102}[f(\theta)]\mathbf{D}(\pi)\mathbf{H}_{17,102}^{\mathrm{T}}$ correspondingly.

The reconstructed $N = 102$-variate Toeplitz matrices are profoundly different from the true sinc($W$) matrices. In Fig. 31- Fig. 35 for $W = 0.2 - 0.4$ we compare the matrix $\mathbf{T}_N(K)$ eigenvalues of the reconstructed $N = 102$-element matrices with the true eigenvalues of matrix $\mathbf{T}_N(W)$.

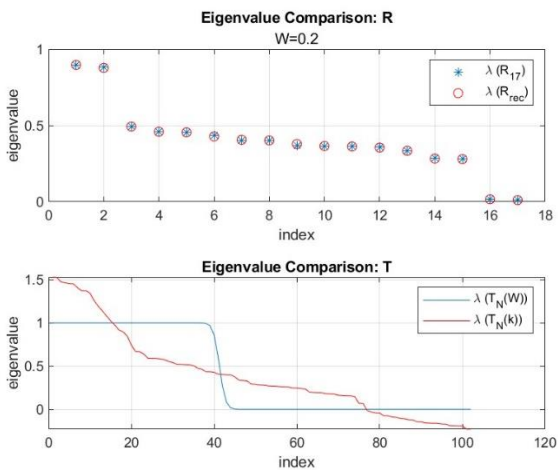

Fig. 31. Eigenvalues: MRA: True ($R_{17}$) vs. reconstructed ($R_{rec}$) (top). True ($T_N(W)$) vs reconstructed ($T_N(k)$)) (bottom). W=0.2.

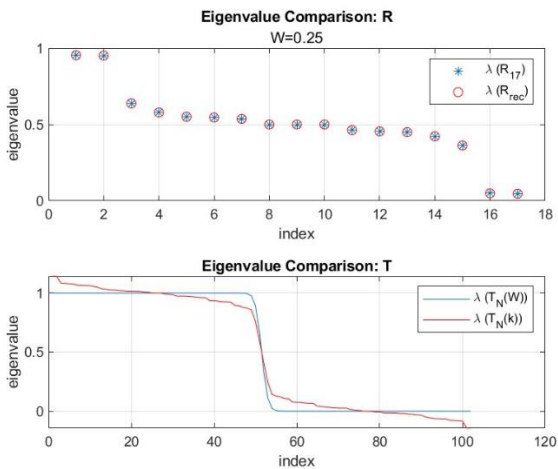

Fig. 32. Eigenvalues: MRA: True ($R_{17}$) vs. reconstructed ($R_{rec}$) (top). True ($T_N(W)$) vs reconstructed ($T_N(k)$)) (bottom). W=0.25.

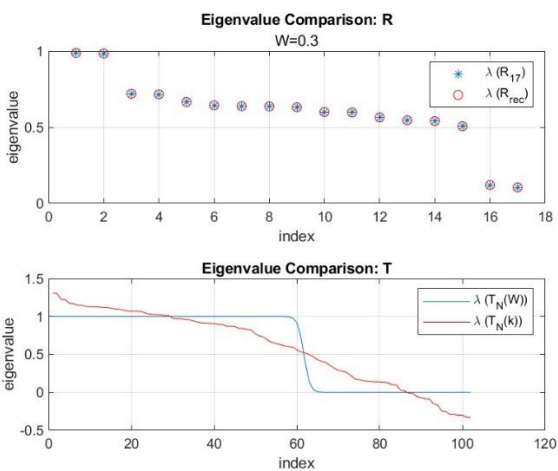

Fig. 33. Eigenvalues: MRA: True ($R_{17}$) vs. reconstructed ($R_{rec}$) (top). True ($T_N(W)$) vs reconstructed ($T_N(k)$)) (bottom). W=0.3.

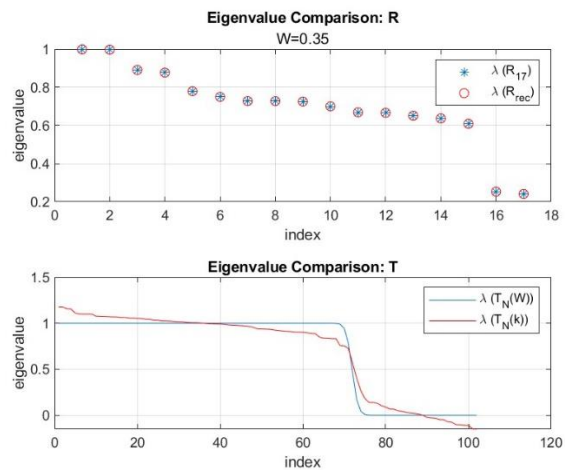

Fig. 34. Eigenvalues: MRA: True ($R_{17}$) vs. reconstructed ($R_{rec}$) (top). True ($T_N(W)$) vs reconstructed ($T_N(k)$)) (bottom). W=0.35.

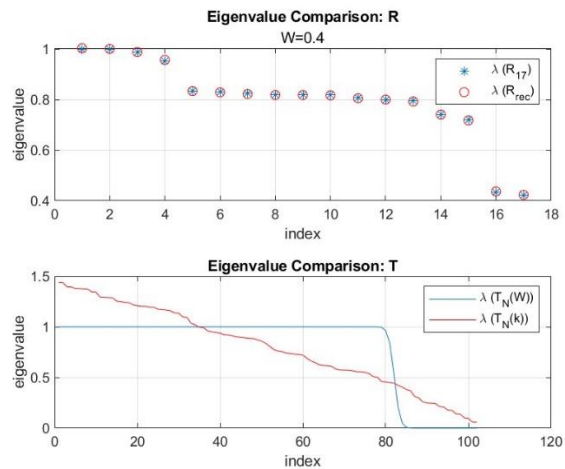

Fig. 35. Eigenvalues: MRA: True ($R_{17}$) vs. reconstructed ($R_{rec}$) (top). True ($T_N(W)$) vs reconstructed ($T_N(k)$)) (bottom). W=0.4.

The figures demonstrate no resemblance. Therefore, similar to the complex-valued case, we have to conclude that the moduli of all $N$ symmetric Toeplitz matrix elements and $M \ll N$ eigenvalues of its MRA transform:

$$\mathbf{H}_{M,N} \mathbf{T}_N \mathbf{H}_{M,N}^{\mathsf{T}} \tag{100}$$

are insufficient for uniqueness of the $M$-variate symmetric MRA matrix $\mathbf{R}_N$ and the $N$-variate Toeplitz symmetric matrix reconstruction. Note that the number of large eigenvalues (equal to 1) of the $N$-variate Toeplitz matrix in all examples above exceeded the MRA dimension. Indeed, the number of such eigenvalues in the sinc($W$) matrix is equal to [36]

$$2WN \gg 17 \tag{101}$$

even for the smallest analyzed $W = 0.2$.

Let us now consider the case, where

$$\mathbf{T}_N = \alpha \mathbf{I}_N + \text{sinc}(0.05)\,, \tag{102}$$

where the number of eigenvalues equal to $(1 + \alpha)$ is $10.2 < 17$. Therefore, the criterion (108) regarding 17 eigenvalues of the matrix $\mathbf{R}_{17}$ may now be augmented by $102 - 17 = 85$ additional requirements on the last eigenvalues of the matrix $\mathbf{T}_N[f(\theta)]$ to be equal to $\lambda_{min} = \alpha$. Computation results for $\alpha = 0.01$ are illustrated by TABLE VI. and Fig. 36.

TABLE VI.

| $\lambda$ ($R_{17}$) | $\lambda$ (restored) | Error |
|---|---|---|
| 0.374046115 | 0.374046115 | 2.77556E-16 |
| 0.316442206 | 0.316442206 | 5.55112E-17 |
| 0.191959795 | 0.191959795 | 1.38778E-16 |
| 0.168504621 | 0.168504621 | 1.11022E-16 |
| 0.140356603 | 0.140356603 | 5.55112E-17 |
| 0.093741231 | 0.093741231 | 2.77556E-17 |
| 0.091921618 | 0.091921618 | 6.93889E-17 |
| 0.091909122 | 0.091909122 | 8.32667E-17 |
| 0.091908772 | 0.091908772 | 4.16334E-17 |
| 0.090763085 | 0.090763085 | 9.71445E-17 |
| 0.047043232 | 0.047043232 | 2.77556E-17 |
| 0.011848241 | 0.011848241 | 2.77556E-17 |
| 0.002419776 | 0.002419776 | 1.51788E-17 |
| 0.001131301 | 0.001131301 | 1.38778E-17 |
| 0.001004256 | 0.001004256 | 2.40693E-17 |
| 0.001000025 | 0.001000025 | 1.0842E-17 |
| 0.001000001 | 0.001000001 | 3.46945E-17 |

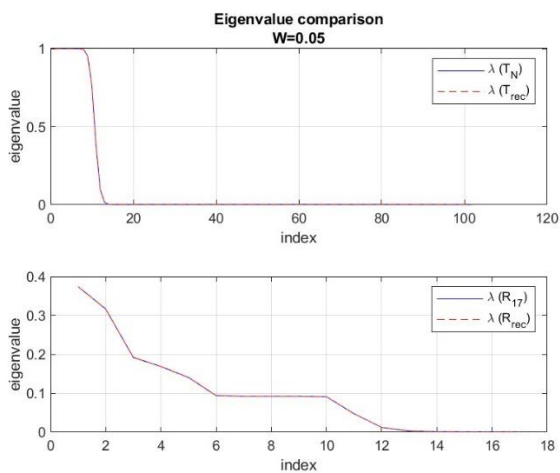

Fig. 36. Eigenvalues: True ($T_N$, N=102) vs reconstructed ($T_{rec}$) (top). MRA: True ($R_{17}$) vs. reconstructed ($R_{rec}$) (bottom).

This data shows that for the case with the signal subspace dimension limited to the MRA array dimension, the reconstruction of the $N$-variate symmetric Toeplitz matrix may be achieved. The same property should embrace the Hermitian matrix case with the properly formulated optimization problem similar to (67)-(76).

## VI. SELECTION OF THE APPROPRIATE ISOMORPHIC SOLUTION FOR ANTENNA CALIBRATION

As demonstrated above, as a result of the symmetric Toeplitz (real-valued) matrix reconstruction, one of the two isomorphic solutions may be obtained. Yet, the appropriate reconstruction of the real-valued symmetric matrix should have the angular spectrum symmetric with respect to the boresight $\theta = 0^o$ and be localized within the visible sector $-\pi/2 \leq \theta \leq \pi/2$. The isomorphic solution is "physical" only if the ULA array is under-sampled, i.e. $d/\lambda \geq 0.5$, and when $\theta_o$ is the solution of the equation:

$$\frac{2\pi d}{\lambda} \sin \theta_o = \pi ; \quad |\sin \theta_o| = \frac{\lambda}{2d} \leq 1 \quad (103)$$

For "over-sampled" operations, when

$$\frac{d}{\lambda} < 0.5 , \quad (104)$$

this solution belongs to the "invisible" domain [37].

As mentioned in Introduction, this study is motivated by the HF OTHR receive antenna array calibration problem. Since these arrays operate within the entire HF band (5 MHz – 35 MHz), the inter-element distance $d$ in the receive arrays is selected to avoid grating lobes within the array coverage for the entire operational frequency band. This means that for most of the HF frequencies, the Rx ULA operates in the "over-sampled" regime (103). For $d/\lambda < 0.5$, the visible angular sector $|\theta_o| \leq \pi/2$, is therefore localized within

$$\mu \equiv \sin \theta ; \quad |\sin \theta| \leq 1 , \quad |\mu| < \frac{2\pi d}{\lambda} < \pi \quad (105)$$

For the "over-sampled" regime, the inappropriate isomorphic solution, on the contrary, should be heavily presented in the "invisible" sector

$$\left| \frac{2\pi d}{\lambda} \right| \leq |\mu| \leq \pi \quad (106)$$

For the small portion of the highest HF frequencies, the HF OTHR Rx array may operate in the under-sampled regime. Yet, even in this case the spatial spectrum would be shifted to the angle $\theta_o = \arcsin((1/2)(\lambda/d))$, which should be far away from the boresight direction $\theta = 0^o$. Moreover, for a broad enough angular spectrum it should be heavily presented at both ends of the visible arc, which is highly unlikely for most practical signals (clutter returns). Therefore, for both $d/\lambda < 1/2$ and $d/\lambda \geq 1/2$, it should not be a problem to identify the spatial spectrum symmetric with respect to $\theta_o = 0^o$ against the one shifted into the invisible or visible domain.

In order to perform the selection of the two isomorphic solutions, the Maximum Entropy (ME) spectrum $\mathbf{S}(\mu)$ should be calculated:

$$\mathbf{S}(\mu) = \frac{1}{\left| \alpha_\mu^H \mathbf{T}_N^{-1} \mathbf{e}_1 \right|^2} \quad (107)$$

$$-1 \leq \mu \leq 1$$
$$\mathbf{e}_1^T = (1, 0, \dots, 0)$$

Recall that for $d/\lambda < 1/2$, the visible sector is limited by

$$|\mu| < \frac{2\pi d}{\lambda} \quad,$$

while the range $2\pi d/\lambda \leq |\mu| \leq \pi$ corresponds to the invisible domain. For the ULAs with $d/\lambda > 1/2$, the region

$$\pi \leq \mu < \frac{2\pi d}{\lambda} \tag{108}$$

corresponds to the grating lobes area.

In Fig.37-Fig.38, we introduce the ME spectra calculated for an $N = 102$-element ULA for the sinc($W$) covariance matrix with $W = 0.2$ - $0.4$ and both isomorphic solutions. These figures support our observation of the straightforward ability to select the solution with the ME spectrum symmetric with respect to $\theta_o = 0^o$.

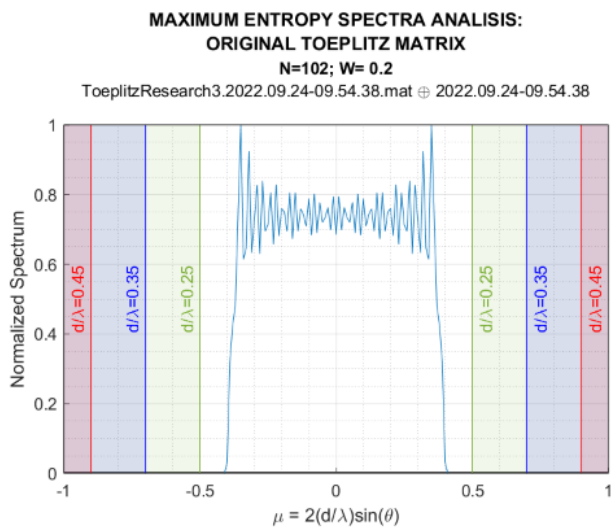

Fig.37. Maximum Entropy Spectra. Original Toeplitz Matrix

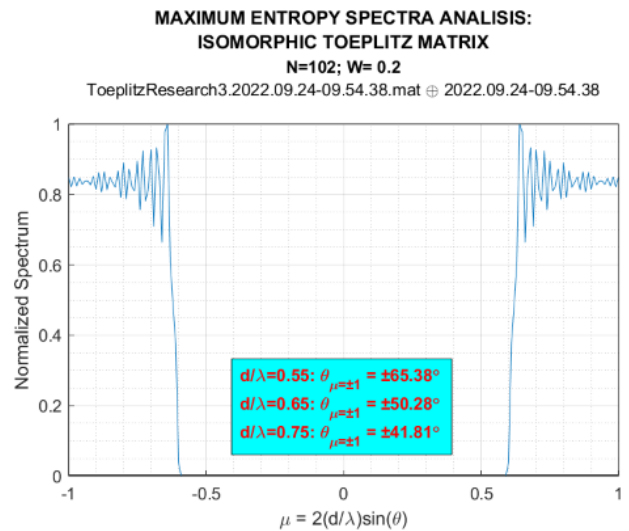

Fig.38. Maximum Entropy Spectra. Isomorphic Toeplitz Matrix

For the properly identified isomorphic solution, the calibration phase errors $(\gamma_n - \gamma_o)$, $n = 1, \dots, N-1$ may be estimated as

$$(\gamma_n - \gamma_o) = \arg[\mathbf{R}_N(f(\theta))]_{1n} - \arg[\mathbf{T}_N(f(\theta))]_{1-n} \tag{109}$$

$$n = 1, \dots, N-1$$

where

$$\arg[\mathbf{T}_N(f(\theta))] = \begin{cases} \pi \\ 0 \end{cases} \tag{110}$$

depending on the sign of the sub-diagonal in the reconstructed real-valued Toeplitz matrix.

For the $M$-element MRA array, the procedure is very similar. After successful restoration of the $N$-valued symmetric Toeplitz matrix with the proper isomorphic solution selected, we construct the $M$-variate MRA covariance matrix transformation $\mathbf{H}_{M_1N}\mathbf{T}_N\mathbf{H}_{M_1N}^T$. Comparison of the Hermitian matrix $\mathbf{H}_{M_1N}\mathbf{T}_N\mathbf{H}_{M_1N}^T$ with the actual MRA covariance matrix elements should provide the estimates of the "calibration" phases.

## VII. Impact of a finite sample support on the Toeplitz matrix reconstruction methodology

In practical applications, the precise covariance matrices are not known and are usually replaced by the sample covariance matrices or their "robust" derivatives. For the (complex) Gaussian data, the standard sample matrix constructed by $T \geq N$ i.i.d. training samples $\mathbf{X}_t \sim \mathcal{CN}(0, \mathbf{R}_N)$ is the maximum likelihood estimate. The rigorous ML estimate of the Toeplitz covariance matrix is not known so far, with a number of practical quasi-optimal routines being proposed [38]-[40]. One of the simplest routines is "redundancy averaging" [39], [40], which is the construction of the Toeplitz matrix by averaging over the diagonals of the ML Hermitian matrix estimate. This method produces minimal errors for the estimated covariance matrix lags, but does not guarantee the positive definiteness of the "redundancy averaged" matrix. An alternative averaging approach that guarantees the positive definiteness creates a biased estimate and is also inappropriate in some applications.

In this regard, it is important to specify the type of convergence required for the successful solution of the "Toeplitz inverse eigenvalue problem", i.e. the Toeplitz matrix reconstruction, and ultimately, for proper ULA calibration. In signal processing applications associated with the covariance matrix estimation, one usually discriminates the "weak" or "the criterion" convergence against the "strong" or "argumental" convergence [27]-[29]. The "weak" convergence is associated with the specific criterion the covariance matrix is estimated for. For example, for interference mitigation with a few interfering point sources, the number of independent training samples, equal to the number of point sources, accurately form the linear subspace occupied by the interferers. Therefore, designing a filter (antenna) orthogonal to this subspace provides sufficient, in most cases, interference rejection. On the contrary, the "strong" convergence is evaluated by the errors of the estimated covariance matrix elements. While the "weak" convergence could be accelerated depending on the criterion, the "strong" convergence depends on the number of i.i.d. training samples and cannot be accelerated [27]-[29].

For the Toeplitz matrix reconstruction problem, the rank of the matrix is irrelevant, and the accuracy of the reconstruction is evaluated with the magnitude errors that the elements of the covariance matrix are estimated with. Finally the same estimation errors that specify the "calibration" phase errors are estimated. This discussion demonstrates that since we are interested in the "strong" or "argumental" convergence of the reconstructed Toeplitz matrix, we should expect a significant number of i.i.d training samples required for an accurate enough reconstruction.

Indeed, the "redundancy averaging" over the sample covariance matrix $\widehat{\mathbf{R}}_N$ elements (in absence of calibration errors) leads to the following "argumental" convergence [44]:

$$P \left\| \widehat{\mathbf{T}}_N^{(J)} - \mathbf{T}_N \right\| > x$$
$$= \exp \left[ -\frac{cTx^2}{4M_f \log T} (1 + O(1)) \right] \tag{111}$$

where $T \to \infty, N/T \to C$, and $t_n$ are absolutely summable with $t_o \neq 0$. In (111), $M_f = \text{essup } f(\omega)$ is the essential supremum of the covariance matrix $\mathbf{T}_N$ power spectrum $f(\omega)$ [45]:

$$\mathbf{T}_N(f) = \left[ \frac{1}{2\pi} \int_0^\infty f(\omega) e^{-i(k-j)\omega)} d\omega \right],$$
$$k, j = 0, \dots, N-1 \tag{112}$$

and

$$\|\mathbf{A}\| = \max \frac{|\mathbf{A}x|_2}{|x|_2} \tag{113}$$

is the matrix $\mathbf{A}$ spectral norm. One can see that this "strong" convergence rate

$$\frac{cT}{\log T} \to \frac{N}{\log T} \tag{114}$$

is indeed very slow.

While the large required training sample volume support is a bit disappointing, the "argumental" type of the required convergence allows for any consistent estimation technique to be applied. For this reason, we next compare the eigenvalues of the reconstructed real-valued Toeplitz matrix with the eigenvalues of the original sample matrix and the eigenvalues of the "redundancy averaged" Toeplitz matrix. When directly comparing the elements of the reconstructed and true Toeplitz matrix, we should consider the significant dynamic range of the covariance lags with some of them reaching practically zero level. Therefore, the relative errors should be minimal for the comparatively large covariance lags, both in the sample matrix and "redundancy averaged" one.

Let us specify our first algorithm, when dealing with the sample matrix $\widehat{\mathbf{R}}_N$. For the reconstruction, we average the elements moduli over the sub-diagonals of the sample matrix and use this positive-valued Toeplitz matrix as the initial matrix for the sign inversion distribution. The Toeplitz matrix that we reconstruct aims to replicate the eigenvalues of the sample covariance matrix as accurately as possible. In Monte-Carlo simulations, we compare the true covariance matrix $\text{sinc}_N(W)$ elements and its eigenvalues with the similar properties of the sample matrix $\widehat{\mathbf{R}}_N$, the properties of the "redundancy averaged" sample matrix $\widehat{\mathbf{R}}_{rA}$ (with the "calibration" phase errors absent), and the elements and eigenvalues of the reconstructed Toeplitz matrix.

For simulations, we selected the $N = 102$-element ULA with $d/\lambda = 1/2$ and rectangular angular spectrum shifted in coning angle with respect to the boresight by $\theta = 20^o$. The covariance matrix of the signals at the output of the antenna array is:

$$\mathbf{R}_N = \alpha \mathbf{I}_N + \mathbf{\Phi}_N \mathbf{T}_N(W) \mathbf{\Phi}_N^{\text{H}} \tag{115}$$

where $\alpha$ is the input noise-to-signal ratio (SNR$^{-1}$) and $W = 0.15$ in the $\text{sinc}_{102}(W)$ matrix. For $\alpha = 0.01$ and $W = 0.15$, the input SNR is SNR $= 2W/\alpha = 30$, which is a typical SNR for real-valued signals used in testing. The matrix $\mathbf{\Phi}_N$ is the diagonal matrix

$$\mathbf{\Phi}_N = \text{diag} \left[ \exp j \left( \varphi_n + \frac{2\pi d}{\lambda} n \sin \theta_o \right) \right],$$
$$n = 0, \dots, 101 \tag{116}$$

consisting of the random calibration errors $\varphi_n$ and regular angular shift $\theta_o = 20^o$. The random phases $\varphi_n, n = 1, \dots, N-1$ were generated as uniformly distributed random numbers over the interval $[-\pi, \pi]$. Finally, the sample matrix $\widehat{\mathbf{R}}_N$ was generated as:

$$\widehat{\mathbf{R}}_N = \frac{1}{T} \sum_{t=1}^T \mathbf{X}_t \mathbf{X}_t^{\text{H}}, \quad \mathbf{X}_t \sim C \, \mathcal{N}(0, \mathbf{R}_N) \tag{117}$$

$$X_t = R_N^{\frac{1}{2}} \xi_t, \quad \xi_t \sim \mathcal{CN}(0, I_N) \qquad (118)$$

Note that for the sample matrix $\hat{R}_N(T)$, $T \geq N$, the complex Wishart distribution is available, as well as the distribution of its eigenvalues [41].

In Fig. 39 we introduce the eigenvalues of the true covariance matrix $R_N$ (interdependent of $\Phi_N$), the eigenvalues of the sample matrix $\hat{R}_N$ for different $T$, and the eigenvalues of the reconstructed Toeplitz matrix. Also for comparison, we introduce the eigenvalues of the redundancy averaged (no calibration phase errors) sample matrix $\hat{R}_{r_A}(T)$. The figures are calculated for $T = 60$, $306$, $3000$, and $3x10^4$.

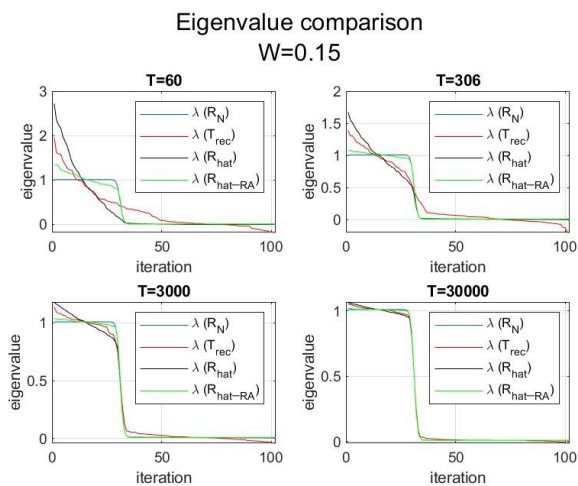

Fig. 39. Eigenvalues (no phase errors): $R_N$ = true covariance matrix. $T_{rec}$ = reconstructed. $R_{hat}$ = sample matrix ($\hat{R}_N$). $R_{hat\_RA} = \overline{\hat{R}_N}$ with redundancy averaging ($\hat{R}_{r_A}(T)$)

Analysis of these results proved our expectations regarding the "argumental" ("strong") nature of convergence. Indeed, we can see that the "stair-like" nature of the $sinc_N(W)$ matrix eigenvalues for $W = 0.15$ is not accurately replicated by the sample matrix eigenvalues up to $T \geq 3000$ ($T \geq 30N$). For $T = 60 - 3000$, the eigenvalues of the sample redundancy averaged and reconstructed Toeplitz matrices have little resemblance with the eigenvalues of the true covariance matrix. Redundancy averaging (in absence of "calibration" errors) somewhat improved this resemblance, but for $N > T = 60$ it created a few negative eigenvalues for the smallest values.

The most surprising aspect is the very accurate restoration of the sample matrix eigenvalues in the reconstructed Toeplitz matrix. Note that the sample matrix $\hat{R}_N$ is not Toeplitz and does not have the same moduli over the matrix sub-diagonal elements. And yet, the reconstructed Toeplitz matrix very accurately restored the eigenvalues of the Hermitian sample matrix by distributing the sign inversions over the sub-diagonals with the averaged moduli.

Let us now analyze the element-wise ("argumental") convergence of the reconstructed Toeplitz matrices. In TABLE VII. we introduce the first column elements of the accurate matrix $sinc_N(W)$ for $N = 102$, $W = 0.15$, and the first column of the reconstructed Toeplitz matrix with redundancy averaging

(no calibration phase errors). As expected, the magnitude of the true Toeplitz matrix elements ranges from 0.31 to $10^{-17}$.

TABLE VII.

| index | $|sinc_N(W)|$ | $|reconstructed|$ | 0=same sign, 1=different sign |
|---|---|---|---|
| 1 | 0.31 | 0.310726091 | 0 |
| 2 | 0.257518107 | 0.258225412 | 0 |
| 3 | 0.151365346 | 0.152040669 | 0 |
| 4 | 0.032787721 | 0.033456735 | 0 |
| 5 | -0.046774464 | -0.046143831 | 0 |
| 6 | -0.063661977 | -0.063038117 | 0 |
| 7 | -0.031182976 | -0.030524709 | 0 |
| 8 | 0.014051881 | 0.014587966 | 0 |
| 9 | 0.037841336 | 0.038102214 | 0 |
| 10 | 0.028613123 | 0.028508514 | 0 |
| 11 | 1.16945E-17 | 0.00045792 | 0 |
| 12 | -0.023410737 | -0.024195162 | 0 |
| 13 | -0.025227558 | -0.026157214 | 0 |
| ... | ... | ... | ... |
| 93 | -0.003290551 | -0.002132311 | 0 |
| 94 | -0.001057668 | -0.000587597 | 0 |
| 95 | 0.001990403 | 0.003627571 | 0 |
| 96 | 0.00335063 | 0.004508928 | 0 |
| 97 | 0.001948936 | 0.002842299 | 0 |
| 98 | -0.001014053 | -0.000180602 | 0 |
| 99 | -0.003089089 | -0.000861365 | 0 |
| 100 | -0.002601193 | -0.000995127 | 0 |
| 101 | -3.43118E-17 | 0.005627453 | 1 |
| 102 | 0.002549684 | 0.00953267 | 0 |

Yet, the relatively strong elements with $|t_n| = 0.31 - 0.02$ are estimated with reasonable accuracy for $T = 3000$. One can see that in this case up to 13 first sub-diagonals estimates have the same sign and are close to the true magnitude. For $T = 3x10^2$ and $3x10^4$, this correspondence extends over the greater number of sub-diagonals. Yet, even for $T = 3x10^4$, the smallest covariance lags are erroneously estimated, with the erroneous sign allocation.

The correct sign allocation in the reconstructed Toeplitz matrix is important for the phase errors estimation. Indeed, if the elements of the reconstructed Toeplitz matrix are accurate, then we have:

$$\arg R_{pl} - \arg T_{p-l} = \varphi_p - \varphi_l - \frac{2\pi d}{\lambda}(p - l)\sin\theta_o \qquad (119)$$

For very small moduli the sign allocation may be wrong and may require a correction.

Let $\widehat{\mathbf{R}}_N$ be the sample covariance matrix with averaged moduli over the sub-diagonals, and let $\mathbf{T}_N$ be the reconstructed Toeplitz matrix. Then for the properly assigned sign inversions the result of the element-wise (Hadamard) division

$$\widehat{\mathbf{R}}_N \oslash \mathbf{T}_N = \mathbf{\Phi}_N \mathbf{1}_N \mathbf{1}_N^{\mathrm{T}} \mathbf{\Phi}_N^{\mathrm{H}} \qquad (120)$$

should tend to the matrix of the first rank with the eigenvector corresponding to the maximum eigenvalue equal to

$$\mathbf{V}_1^{\mathrm{T}} = \mathrm{const}[1, \exp\left(j\varphi_1 + \frac{2\pi d}{\lambda}\sin\theta\right), \dots, \\ \exp\left(j\varphi_{N-1} + \frac{2\pi d}{\lambda}(N-1)\sin\theta\right)] \qquad (121)$$

There are two important advantages of the approach associated with the reconstruction of the Toeplitz matrix $\mathbf{T}_N$ in (121) with the known elements moduli, such that the Hadamard division in (120) is a rank-one matrix. The first is this approach may lead to a solution that does not depend on the number of close to zero moduli. The second is the rank-one property is retained for any $M$-variate sub-matrix of the matrix $\mathbf{R}_N$ in (121). This leads to a possible iterative solution of this problem.

Specifically, one can solve a sequence of problems with the increasing dimension $M \to N$:

Find $\sigma_{l-p} = \begin{cases} 1 \\ -1 \end{cases}$ $(l - p \neq 0)$ such that

$$\mathrm{rank}\left[\mathbf{A}_M = \mathbf{\Phi}_M \mathbf{T}_M \mathbf{\Phi}_M^{\mathrm{H}} \otimes \left[\frac{\sigma_{l-p}}{|t_{l-p}|}\right]\right] = 1 \\ l, p = 1, \dots M \qquad (122)$$

Obviously, (122) is equivalent to

Find $\sigma_{l-p} = \begin{cases} 1 \\ -1 \end{cases}$ $(l - p \neq 0)$, for

$$\mathrm{rank}\ \mathbf{A}_M = 1 \qquad (123)$$

$$\mathbf{A}_M = \left[\frac{\mathrm{R}_{pl}}{|\mathrm{R}_{pl}|}\right] \otimes \|\sigma_{p-l}\|, \quad p, l = 1, \dots, M \qquad (124)$$

Note that the reconstruction of the bi-phase Toeplitz matrix $\|\sigma_{p-l}\|$ in (123) may lead to one of the two isomorphic solutions:

$$\mathbf{A}_M = \mathbf{\Phi}_M \mathbf{1}_M \mathbf{1}_M^{\mathrm{T}} \mathbf{\Phi}_M^{\mathrm{H}} \qquad (125)$$

or

$$\mathbf{A}_M = \mathbf{\Phi}_M \ \mathrm{diag}\ [e^{j\pi m}] \mathbf{1}_M \mathbf{1}_M^{\mathrm{T}} \ \mathrm{diag}\ [e^{-j\pi m}] \\ m = 0, \dots, M-1 \qquad (126)$$

Obviously, this is the same "stability group" problem specified above by the M.T. Chu theorem 2.3. The resolution of this ambiguity may be performed as per Sec VI, by the analysis of the maximum entropy spectrum of the "calibrated" covariance matrix:

$$\mathbf{\Phi}_N^{\mathrm{H}} \mathbf{R}_N \mathbf{\Phi}_N = \begin{cases} \mathbf{T}_N \ , \\ \mathrm{diag}\ [e^{-j\pi n}] \mathbf{T}_N \ \mathrm{diag}\ [e^{j\pi n}] \ , \\ n = 0, \dots, N-1 \end{cases} \qquad (127)$$

with the "non-physical" solution (for the over-sampled ULAs heavily residing in the invisible domain. For this reason, moving forward we mostly report on the accuracy of the proper ("physical") solution.

To complete our discussion of the diagonal matrix $\mathbf{\Phi}_N$ estimation for the true (accurate) matrix $\mathbf{R}_N$ in (114), we have to specify our iterative technique with the progression $M \to N$. It is important that there are various ways of such progression. Suppose that by the "dynamic programming" technique we accurately solved the $N > M$-variate problem and got the binary Toeplitz matrix $\|\sigma_{p-l}\|, p, l = 1, \dots, M$, that converts the matrix $\mathbf{A}_M$ in (122) into a rank-one matrix. Then for the $(M + 1)$-dimension, we have to specify only a single "corner" element in the Toeplitz binary matrix $\|\sigma_{p-l}\|, p, l = 1, \dots, M + 1$, that should convert the $(M + 1)$-variate matrix $\mathbf{A}_{M+1}$ into a rank-one matrix. Obviously, optimization of this "corner" element means the selection of one of the two possible values [+1, -1]. Instead of increasing the dimension by one, one can keep the number of sub-diagonals, optimized by the "dynamic programming" constant, moving from $\mathbf{A}_M$ to $\mathbf{A}_{2M}$ and so on, until the full dimension $N$ is reached. This technique was used in the Monte-Carlo simulations discussed below.

Finally, one may keep the decomposition of the $N$-variate problem as a number of $M$-variate problems. Indeed, if the two consecutive $M$-variate (central) matrices have at least one common element, then the partial $M$-variate solutions for the vectors in $\mathbf{\Phi}_n$ may be composed of one $N$-variate vector for $\mathbf{\Phi}_N$. Obviously, if they have up to $(M - 1)$ common elements ("sliding by one element along the array"), then the solutions for these common elements may be averaged, if applied to the sample matrix instead of the true ones. For the accurate (true) matrices $\mathbf{R}_N$ and $\mathbf{T}_N$ in (120), all these techniques lead to the same results.

Let us now concentrate on the practical case, when the covariance matrix $\mathbf{R}_N$ is represented by the sample matrix $\widehat{\mathbf{R}}_N$:

$$\widehat{\mathbf{R}}_N = \mathbf{\Phi}_N \widehat{\mathbf{R}}_N^{\mathrm{T}} \mathbf{\Phi}_N^{\mathrm{H}} ; \\ \widehat{\mathbf{R}}_N^{\mathrm{T}} = \frac{1}{T}\sum_{t=1}^{T} \mathbf{X}_t \mathbf{X}_t^{\mathrm{H}}, \quad \mathbf{X}_t \sim \mathcal{CN}(0, \mathbf{T}_N(W)) \qquad (128)$$

The sample matrix $\widehat{\mathbf{R}}_N^{\mathrm{T}}$ is not a Toeplitz matrix for any finite $T$ in (128), and therefore, the estimation algorithm should be optimized to provide the minimum estimation error. Strictly speaking, the problem of joint maximum likelihood estimation of the real-valued Toeplitz matrix $\mathbf{T}_N$ and the diagonal matrix $\mathbf{\Phi}_N$ has to be formulated for the model (128). With the well-known problem of the rigorous ML estimation of the Toeplitz matrix alone, the problem of the joint ML estimation of $\mathbf{T}_N$ and $\mathbf{\Phi}_N$ is not going to be any simpler, and therefore it deserves a separate investigation. In particular, the techniques of the problem solution mentioned above have to be analyzed since they may demonstrate different accuracies. Yet, as discussed above, the "argumental" (strong) nature of the required convergence suggests that one may expect quite a modest accuracy improvement for $\widehat{\boldsymbol{\Phi}}_N$ by the application of the optimal or quasi-optimal algorithm compared to any consistent technique discussed above.

Finally note that for the Hermitian Toeplitz matrix case we may apply a similar approach, constructing the Toeplitz matrix $\|\exp(j\sigma_{p-l})\|$, $p, l = 1, \ldots, N$ that converts the $N$-variate matrix $\left\|[\widehat{\mathbf{R}}_N]_{pl}/|\widehat{\mathbf{R}}_{Npl}|\right\|$ $p, l = 1, \ldots N$ into the closest to rank-one problem. Instead of the integer problem:

find

$$\|\sigma_{p-l}\|, \quad p, l = 1, \ldots N$$
$$\sigma_{p-l} = \begin{cases} +1, \\ -1. \end{cases} \quad p \neq l \qquad (129)$$

such that

$$\sum_{i=2}^{N} \left| \lambda_i \left[ \frac{[\widehat{\mathbf{R}}_N]_{pl}}{|\widehat{\mathbf{R}}_{Npl}|} \otimes \sigma_{p-l} \right] \right| = \min \qquad (130)$$

where one has to solve the ToIEP problem:

find

$$\|\exp(j\phi_{p-l})\|, \quad p, l = 1, \ldots, N, \qquad (131)$$

such that

$$\sum_{i=2}^{N} \left| \lambda_i \left[ \frac{[\widehat{\mathbf{R}}_N]_{pl}}{|\widehat{\mathbf{R}}_{Npl}|} \otimes \exp(j\phi_{p-l}) \right] \right| = \min \qquad (132)$$

and then applying the Newton's technique, as per Sec. III.

Let us now analyze the results of Monte-Carlo simulations, starting from the case with the accurate (true) matrix $\mathbf{R}_N$ in (120) followed by the cases with the finite sample support $T = 3x10^2 - 3x10^5$ for $N = 102$. For the true covariance matrix (infinite sample volume) in (120) we provide an example calculated for $W = 0.2$, $N = 102$, with the true estimated phases, the phases in the two solutions (125), (126), and the difference between them. This example illustrates the case where the isomorphic solution $\mathbf{D}(\pi)\mathbf{D}(\boldsymbol{\Phi}_N)\mathbf{D}(\pi)$ is the correct solution.

TABLE VIII.

| index | $D(\boldsymbol{\Phi}_N)$ (est. phase) | true phase | $D(\boldsymbol{\Phi}_N)$– true phase | $D(\pi)D(\boldsymbol{\Phi}_N)D(\pi)$– true phase |
|---|---|---|---|---|
| 1 | 0 | 0 | 0 | 0 |
| 2 | 5.105524352 | 1.963931698 | 3.14 | 0 |
| 3 | 0.462555794 | 0.462555794 | 0 | 0 |
| 4 | -3.740603352 | -0.599010698 | -3.14 | 0 |
| 5 | -0.672584415 | -0.672584415 | 0 | 0 |
| 6 | -3.905487812 | -0.763895159 | -3.14 | 0 |
| 7 | 2.326057755 | 2.326057755 | 0 | 0 |

| index | $D(\boldsymbol{\Phi}_N)$ (est. phase) | true phase | $D(\boldsymbol{\Phi}_N)$– true phase | $D(\pi)D(\boldsymbol{\Phi}_N)D(\pi)$– true phase |
|---|---|---|---|---|
| 8 | 5.200772704 | 2.05918005 | 3.14 | 0 |
| 9 | 2.898839739 | 2.898839739 | 0 | 0 |
| 10 | 4.259334622 | 1.117741969 | 3.14 | 0 |
| ... | ... | ... | ... | ... |
| 93 | 1.150139853 | 1.150139853 | 0 | 0 |
| 94 | 1.984941285 | -1.156651369 | 3.14 | 0 |
| 95 | -0.302938237 | -0.302938237 | 0 | 0 |
| 96 | 0.825103712 | -2.316488941 | 3.14 | 0 |
| 97 | -1.391546843 | -1.391546843 | 0 | 0 |
| 98 | 6.033572355 | 2.891979701 | 3.14 | 0 |
| 99 | 2.201409989 | 2.201409989 | 0 | 0 |
| 100 | 0.478267152 | -2.663325501 | 3.14 | 0 |
| 101 | -2.872068967 | -2.872068967 | 0 | 0 |
| 102 | 5.489133802 | 2.347541149 | 3.14 | 0 |

As follows from the TABLE VIII. one of the two solutions is absolutely accurate for all $W$. The second isomorphic solution has the additional "$\pi$" in every second phase, as expected for the term $\text{diag}[\exp(j\pi n)]$ in (126).

Since the method of the appropriate solution selection is specified, at least for the over-sampled ULAs, in the following we report on the accuracy of the appropriate solution. For every finite $T$ and a given $W$, we conducted ten independent trials with the phase RMSE averaged over all 101 phases ($\varphi_o = 0$) and 10 random sample matrix realizations. The results for $W = 0.1 - 0.45$ and $T = 3x10^2 - 3x10^7$ for $N = 102$-element ULA and $M = 17$, are provided in the TABLE IX.

TABLE IX.

| Average RMSE (deg) for phase estimation of reconstructed Toeplitz real-valued matrix. N=102, M=17 | | | | | |
|---|---|---|---|---|---|
| $W$ | $T=300$ | $T=3e3$ | $T=3e4$ | $T=3e5$ | $T=3e7$ |
| 0.1 | 21.1 | 15.0 | 4.0 | 3.5 | 3.3 |
| 0.15 | 18.9 | 7.6 | 3.2 | 2.4 | 1.7 |
| 0.2 | 25.4 | 8.8 | 5.3 | 4.0 | 4.1 |
| 0.27 | 34.9 | 8.6 | 4.9 | 2.2 | 0.6 |
| 0.3 | 39.8 | 10.8 | 5.4 | 4.6 | 4.4 |
| 0.35 | 63.2 | 11.2 | 5.0 | 3.0 | 1.9 |
| 0.4 | 66.1 | 12.1 | 7.4 | 6.0 | 4.3 |
| 0.45 | 84.1 | 15.0 | 5.5 | 3.2 | 1.9 |

For $T = 3000$, $W = 0.1$, where RMSE = 4.7°, in Fig. 40 we plot the true phases in $\boldsymbol{\Phi}_N$ over the estimated ones from the selected isomorphic solutions.

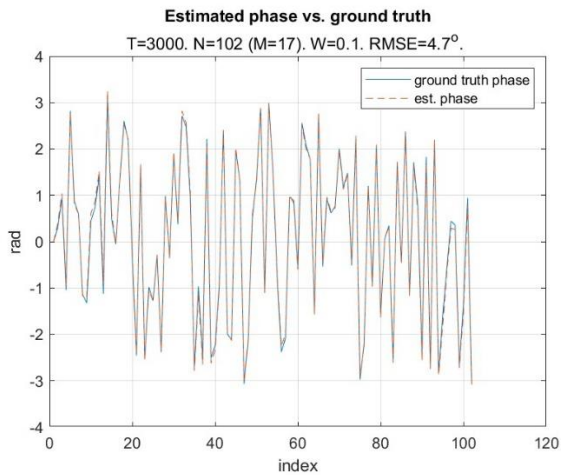

**Estimated phase vs. ground truth**

T=3000, N=102 (M=17), W=0.1, RMSE=4.7°.

Fig. 40. Estimated phase vs. ground truth phase. T=3000, W=0.1.

Recall that the direct reconstruction of the Toeplitz matrix (with the sample matrix eigenvalues) provides a number of poorly estimated phase values due to the wrongly distributed sign inversions over the sub-diagonals with the close to zero moduli. Fig. 40 demonstrates that estimation errors are evenly distributed over the antenna elements in this modified technique.

Finally, the results prove the "argumental" nature of convergence that requires an asymptotically large sample volume, irrespective of the algorithm modifications. In fact, the dependence RMSE $(W, T)$ on $T$ for the given $N = 102$ is quite closely described by (118), with its $N/\log T$ convergence rate. Still, as discussed above, the methodology of joint optimal and/or sub-optimal estimation of the Toeplitz matrix $\mathbf{T}_N$ and the (random) phase errors matrix $\mathbf{\Phi}_N$ deserve a separate investigation. On the other hand, recall that the number of training samples $T = 10^5$ is usually collected by one antenna element receiver over a single coherent integration time interval (CIT, dwell) in HF OTHR, operating with the waveform bandwidth BDW = 40 kHz and CIT = 25 sec.

## VIII. CONCLUSIONS AND RECOMMENDATIONS

In this study, we formulated a number of new problems in the field of the structural (Toeplitz) inverse eigenvalue problems (ToIEP) associated with the phase calibration in uniform linear antenna arrays (ULAs) and suggested techniques for the appropriate modifications. Specifically, since the phase "calibration" errors do not change the eigenvalues and moduli of the original Toeplitz covariance matrix of the phase errors-free ULA antenna array, the formulated ToIEP problems dealt with the reconstruction of the Toeplitz covariance matrix, given its eigenvalues and moduli of all matrix elements. For the fully augmentable Minimum Redundancy Arrays (MRA), the $N$-variate Toeplitz matrix reconstruction should be performed given the moduli of all matrix' elements and eigenvalues of the $M \ll N$-variate covariance matrix of the MRA array.

For the real-valued case, which corresponds to the signal with the symmetric angular spectrum, we confirmed the theorem of M. T. Chu's statement regarding only two available isomorphic solutions to this problem. We demonstrated that for the oversampled regime of ULA operations, typical for HF OTHR receive antenna arrays, the appropriate (one of the two) solution could be easily specified.

For the complex-valued case with Hermitian Toeplitz covariance matrices, the situation was proven to be more complicated. The developed Toeplitz matrix reconstruction technique provided an example for when the reconstructed Hermitian Toeplitz matrix with computationally accurate enough eigenvalues and elements' moduli, did not belong to the anticipated family (8):

$$\mathbf{J}(\mathbf{T}_N) = \text{diag} \left[ \exp(jn\boldsymbol{\varphi}) \right] \mathbf{T}_N \text{diag} \left[ \exp(-jn\boldsymbol{\varphi}) \right] \quad (133)$$
$$n = 0, .., N-1$$

where $\text{diag}[\exp(jn\boldsymbol{\varphi})]$ is the diagonal matrix with the elements $\exp(jn\boldsymbol{\varphi}), n = 0, ..., N-1$ over its main diagonal. The difference between the real-valued and complex-valued cases may be explained by the insufficient number of provided $(2N - 1)$ positive parameters instead of the $(2N - 1)$ real-valued ones required for the Hermitian Toeplitz matrix specification. Failure to produce the solution to the Hermitian Toeplitz matrix reconstruction problem with the given eigenvalues and elements moduli belonging to the family (8), encouraged us to formulate a more stringent ToIEP problem.

Specifically, we formulated the problem of the Hermitian Toeplitz matrix reconstruction with the same elements moduli as the given Hermitian matrix, such that as a result of the element-wise (Hadamard) division of the given Hermitian matrix by the reconstructed Toeplitz matrix, we get a matrix with its rank equal to one, i.e. with all eigenvalues but one equal to zero. The proposed modified Newton's algorithm for all considered cases, including the one with the above mentioned failure, provided solutions that belong to the family (8). Therefore, for both the real-valued symmetric Toeplitz matrices and complex-valued Hermitian Toeplitz matrices, we developed the ToIEP methodology that allows us to decompose the measured covariance matrix at the output of an uncalibrated ULA, as the product of the Toeplitz covariance matrix of the calibrated ULA and diagonal matrices with the phase "calibration" errors.

We demonstrated that the reconstructed Toeplitz covariance matrix may have an arbitrary angular shift of the true angular spectrum of the input signal. For this reason, after restoration of the ULA manifold, the calibrated ULA array may require an additional step of bias compensation. This can be done using weak sources with known coordinates, such as radio stars like Cygnus-A for HF OTHR array calibration. When the phase "calibration" errors are big enough to destroy the desired low sidelobe levels or to secure the low SNR losses in adaptive beamforming but not cause a bias for the stand-alone strong sources, these sources could be used for the bias removal as well.

We demonstrated that, if the signal subspace dimension of the tested signal does not exceed dimensions of a fully augmentable MRA, the same technique may be applied for the MRA array calibration. The algorithm was developed for the real-valued case, but we conjectured the existence of the appropriate algorithm for the complex-valued case of a Hermitian MRA covariance matrix. Monte-Carlo simulations with sample measure covariance matrices estimates in lieu of the accurate covariance matrices demonstrated the anticipated property of the "argumental" or "strong" convergence.

Indeed, for the accurate phase errors estimation by the proposed technique, the elements of the matrix at the output of an uncalibrated ULA should be measured with high precision, regardless of the eigenspectrum of this covariance matrix. As demonstrated by the Monte-Carlo simulations, for an $N = 102$-element ULA, one may require up to $3 \times 10^4$ samples for an accurate enough array calibration. Yet, for an HF OTHR operating with, say 20 kHz bandwidth, over the duration of 1 sec, the required sample volume may be collected.

Obviously, the traditional antenna calibration techniques that use a special set of calibrating test signals should not be disregarded. The proposed technique may be recommended for cases when the traditional calibration techniques are not applicable, such as in the case of passive broadband reception, for example. The example of a successful engineering application of the well-developed mathematical "Inverse eigenvalue problem", provided in this paper, may attract attention to this mathematical instrument for solution of similar signal processing problems.

## ACKNOWLEDGMENT

The authors wish to express their deep gratitude to Prof. M. T. Chu for his guidance and very important advice.

## REFERENCES

[1] S. Friedlander, J. Nocedal, M. L. Overton, "The formulation and analysis of numerical methods for inverse eigenvalue problems", *SIAM J. Numer. Anal.* Vol 24, No 3, June 1987, pp. 634-667.

[2] M.T. Chu, F. Diele, and I. Sgura, "Gradient flow methods for matrix completion with prescribed eigenvalues", *Linear Algebra and its Applications*, vol. 379, pp. 85-112, 2004, doi: 10.1016/S0024-3795(03)00393-8.

[3] Z.-J. Bai, "Inexact Newton methods for Inverse Eigenvalue Problems", Applied Mathematics and Computations, vol. 172, No. 2, 2006, pp. 682-689.

[4] M.T. Chu, "The stability group of symmetric toeplitz matrices, *Linear Algebra and its Applications*, vol. 185, pp. 119-123, 1993, doi: 10.1016/0024-3795(93)90208-6.

[5] F. Noor, S. D. Morgera, "Construction of a Hermitian Toeplitz Matris from an Arbitrary Set of Eigenvalues", *IEEE Trans. on Signal Processing*, vol. 19, No. 8, Aug. 1992, pp. 2093-2094.

[6] M.T. Chu and G.H. Golub, "Inverse eigenvalue problems: theory, algorithms, and applications", *Numerical Mathematics and Scientific Computation*, New York, NY, USA: Oxford University Press, 2005, pp. xviii+387, doi: 10.1093/acprof:oso/9780198566649.001.0001

[7] N. Li, "Some remarks on the inverse eigenvalue problem for real symmetric Toeplitz matrices", ANZIAM. J. 46 (E), 2006, pp. C1327-C1335.

[8] M. Padilla, B. Kolbe, A. Chakrabarty, "The nearest Hermitian Inverse Eigenvalue Problem Solution with Respect to the 2-Norm", arXiv: 1703.00829 v. 1 [math. N2], 2 Marc 2017.

[9] H. J. Landau, "The inverse eigenvalue problem for real symmetric Toeplitz matrices", *Journal of the American Mathematical Society*, vol. 7, No. 3, July 1994, pp. 749-767.

[10] G. Feyh, C. T. Mulls, "Inverse eigenvalue problem for real symmetric Toeplitz matrices", ICASSP-88, 11-14 April 1988, NY, USA, INSPEC Accession Nu.: 3247338, pp. 1636-1639.

[11] W. F. Trench, "Numerical solution of the eigenvalue problem for Hermitian Toeplitz matrices", *SIAM J.* Matrix Anal. Appl. 10 (1985), pp. 135-156.

[12] S. Friedland, "Inverse eigenvalue problems", Linear Algebra and its Applications 17, (1977), pp. 15-51.

[13] M. T. Chu, "Inverse Eigenvalue problems", *SIAM* Rev., Vol. 40, Nu. 1, pp. 1-39, March 1998.

[14] M. T. Chu, G. H. Golub, "Structural inverse eigenvalue problems", *Acta Numerica*, 2002, pp. 1-71.

[15] M. K. Ng, W. F. Trench, "Numerical solution of eigenvalue problem for Hermitian Toeplitz-like matrices", The Australian National University, Technical Report TR-CS-97-14, pp. 1-12

[16] W. F. Trench, "Numerical solution of the inverse eigenvalue problem for real symmetric Toeplitz matrices", *SIAM J.* Sci. Comp. 18 (1997) pp. 1722-1736.

[17] M. T. Chu, "Constructing a Hermitian Matrix from its diagonal entries and eigenvalues", *SIAM Journal on Matrix Analysis and its applications*, Vol. 16, Iss. 4, pp. 1-12.

[18] Z. Liu, L. Chen, Y. Zhang, "The reconstruction of Hermitian Toeplitz matrices with the prescribed eigenpairs", *Journal of System Science and Complexity*, 23, 2010, pp. 961-970.

[19] B. M. Podlevskyi, O. S. Yaroshko, "Newton method for the solution of inverse spectral problems", *Journal of Mathematical Sciences*, Vol. 194, No. 2, October 2013, pp. 156-165.

[20] M. I. Skolnik, G. Nemhaiser, J. W. Sherman, "Dynamic programming applied to unequally spaced arrays", *IEEE Transactions on Antenna and Propagation*, 1964, No. 1, pp. 35-43.

[21] M. Skolnik, G. Nemhauser, L. C. Kefauver, "Thinned, unequally spaced arrays designed by dynamic programming", *Proceedings of the Antennas and Propagation Society International Symposium*, July 1963, pp. 224-227.

[22] A. Paulraj, T. Kailath, "Direction of arrival estimation by eigenstructure methods with unknown sensor gain and phase", Proc IEEE ICASSP'85, Tampa, FL, pp. 640-643, 1985.

[23] A. J. Weiss, B. Friedlander, "Eigenstructure Methods for Direction Finding with Sensor Gain and Phase Uncertainties", *Circuit Systems Signal Process*, Vol. 9, No. 3, 1990, pp.271-300.

[24] J. Yu, J. Krolik, "Adaptive Phase-Array calibration using MIMO Radar clutter", *2013 Radar Conference (RadarCon13)*, Ottawa, ON, Canada, INSPEC Accession Number: 13748146.

[25] A. Weiss, A. Yeredor, "Non-iterative blind calibration of nested arrays with asymptotically optimal weighting", *ICASSP2021*, D01:10.1109 / ICASSP 39728.2021.9415037, pp. 4630-4634.

[26] A. Weiss, A. Yeredor, "Blind calibration of sensor arrays for narrowband signals with asymptotically optimal weighting", *2019 27th European Signal Processing Conference (EUSIPCO)*. 978-9-0827-9703-9/19

[27] A. Weiss, A. Yeredor, "Asymptotically Optimal Blind calibration of Uniform Linear Sensor Arrays for Narrowband Gaussian Signals", *IEEE Trans on Signal Processing*, vol.68, 2020, pp. 5323-5333.

[28] B. T. Polyak, Ya. Z. Tsypkin, "Criterion algorithms of stochastic optimization", Automation and Remote Control (Translation from Russian), vol. 46, No. 6, Part 2, June 1984, pp. 766-771.

[29] B. T. Polyak, Ya. Z. Tsypkin, "Optimal algorithms of criterion optimization under uncertainty", Dokl. Akad. Nauk SSSR, 273, No. 2, 315-318 (1983).

[30] Y. I. Abramovich, "On the "criterion" properties" of adaptive filters optimization problem based on the maximum signal-to-noise criterion", Radio Eng. Elect. Phys. Vol. 26, No. 12, 1981, pp. 74-79 (Translation from Russian).

[31] Palmer, R.D., Vangal S., Larsen, M. F., Fukao S., Nakamura, T., Yamamoto, M. "Phase calibration of VHF spatial interferometry radars using stellar sources", *Radio Sci*., 31, pp. 147-156, 1996.

[32] J. L. Chau, D. L. Hysell, K. M. Kuyeng, F. R. Galindo, "Phase calibration approaches for radar interferometry and imaging configurations: equatorial spread F results", *Ann. Geophys*., 26, 2008, pp. 2333 -2343.

[33] I. S. D. Solomon, D. A. Gray, Y. I. Abramovich, S. Anderson, "Over-the-horizon radar array calibration using echoes from ionized meteor trails", *Radar and Signal Processing, IEE Proceedings*, 145, January 1998, pp. 173-180.

[34] M. T. Chu, Private communication

[35] Y. Abramovich, B. Danilov, "The use of Dynamic Programming for Synthesis of Array Antennas with discrete phase shifters", *Radio Eng. Elect. Phys*., Vol. 21, No. 1, 1976, pp. 69-74.

[36] F. Schwartau, Y. Schröder, L. Wolf, J. Schoebel, "Large Minimum Redundancy Linear Arrays; Systematic Search of Perfect and Optimal Rulers Exploiting Parallel Processing", *IEEE Open, Journal of Antenna and Propagation*, Dec. 2020, PP (99): 1-1, pp. 1-6.

[37] S. Karnik, J. Romberg, M. A. Davenport, "Improved bounds for the eigenvalues of probate spheroidal wave functions and discrete probate spheroidal sequences", arxiv: 2006.00427 v2 [math.CA] 28 Sept. 2020, pp. 1-29.

[38] M. M. Dawoud, A. P. Anderson, "Design of superdirective arrays with high radiation efficiency", *IEEE Trans. on Antenna and Propagation*, Vol. AP-26, No. 6, pp. 819-823, Nov. 1978

[39] J. Burg, "Maximum entropy spectral analysis", PhD Dissertation, Stanford Univ., 1972.

[40] D. A. Linebarger, D. H. Johnson, "The effect of spatial averaging on spatial correlation matrices in the presence of coherent sources", *IEEE Trans. Acoust. Speech, Signal Processing*, Vol. ASSP-38, pp. 880-884, May 1990.

[41] M. Doron, A. Weiss, "Performance of Direction Finding Using Lag Redundancy Averaging", *IEEE Trans. on Signal Processing*, Vol. 41, No. 3, March 1993, pp. 1386-1391.

[42] R. J. Muirhead, Aspects of multivariate statistical theory, John Willey & Sons, Inc., 1982.

[43] Ya. Z. Tsypkin, B. T. Polyak, "Attainable accuracy of adaptation algorithms", *Proceedings of the USSR Academy of Science*, V. 218, No. 3, pp. 532-535 (in Russian)

[44] J. Vinogradova, R. Couillet, W. Hachem, "Estimation of toeplitz covariance matrices in large dimension regime with application to source detection", arXiv: 14031243v1 [csIT], 5 March 2014.

[45] R.M. Gray, "Toeplitz and circular matrices", Now Pub., 2006